\documentclass[english,aps,manuscript]{revtex4}
\usepackage[T1]{fontenc}
\usepackage{textcomp}
\usepackage[latin9]{inputenc}
\usepackage{babel}
\usepackage{array}
\usepackage{pifont}
\usepackage{float}
\usepackage{booktabs}
\usepackage{mathrsfs}
\usepackage{multirow}
\usepackage{varwidth}
\usepackage{amsmath}
\usepackage{amssymb}
\usepackage{graphicx}
\usepackage[a4paper]{geometry}
\usepackage[]
 {hyperref}

\makeatletter

\providecommand{\tabularnewline}{\\}

\@ifundefined{textcolor}{}
{%
 \definecolor{BLACK}{gray}{0}
 \definecolor{WHITE}{gray}{1}
 \definecolor{RED}{rgb}{1,0,0}
 \definecolor{GREEN}{rgb}{0,1,0}
 \definecolor{BLUE}{rgb}{0,0,1}
 \definecolor{CYAN}{cmyk}{1,0,0,0}
 \definecolor{MAGENTA}{cmyk}{0,1,0,0}
 \definecolor{YELLOW}{cmyk}{0,0,1,0}
}

\usepackage{tikz,xcolor,hyperref}

\definecolor{lime}{HTML}{A6CE39}

\AtBeginDocument{
  
}

\makeatother

\begin{document}
\title{\texttt{SCHRÖDINGER OSCILLATOR AND ITS THERMAL PROPERTIES IN A DYNAMICAL
NONCOMMUTATIVE SPACE }}
\author{{\normalsize Ilyas Haouam} \href{https://orcid.org/0000-0001-6127-0408}{\includegraphics{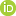}}}
\email{ilyashaouam@live.fr ; ilyas.haouam@usthb.edu.dz}

\address{Theoretical Physics and Didactic Laboratory, Faculty of Physics, University
of Sciences and Technology Houari Boumediene (USTHB), B.P. 32, El
Alia, 16111 Algiers, Algeria. }
\begin{abstract}
In this paper, we study the two-dimensional Schrödinger oscillator
within a dynamical noncommutative (DNC) space. By leveraging perturbation
theory, we derive the energy eigenvalues and eigenvectors and systematically
analyze the effects of both dynamical and non-dynamical noncommutative
settings. First-order corrections to the eigensystem are obtained,
revealing that the energy shift explicitly depends on the DNC parameter
$\tau$. Furthermore, we explore the thermal properties of the system
by employing the partition function. Numerical results are presented
to provide a comprehensive analysis of the system\textquoteright s
behavior under the considered effects. Notably, in the DNC framework,
the commutation relations and the deformation parameter are position-dependent.
Using the two-dimensional Bopp-shift, we effectively map the noncommutative
problem to its commutative counterpart.

{\normalsize$\phantom{}$ }{\normalsize\par}

{\normalsize$\phantom{}$}{\normalsize\par}

\end{abstract}

\maketitle

\section{{\normalsize\label{I}Introduction}}

For decades, the quest for solving wave equations in the presence
of potential energies has remained a central focus in both relativistic
and non-relativistic quantum mechanics \citep{key-1}. These solutions
play a fundamental role in characterizing the physical properties
of various systems, offering deep insights into their underlying behavior.
Furthermore, their significance extends beyond quantum mechanics,
with broad applications in chemical physics and high-energy physics,
particularly in higher-dimensional frameworks \citep{key-2}. Quantum
oscillators, including both the Schrödinger oscillator and the harmonic
oscillator, are foundational systems in quantum mechanics. They are
celebrated for their exact solvability and their extensive applicability
in various physical contexts. Moreover, quantum oscillators are pivotal
in fields such as quantum optics, condensed matter physics, and statistical
mechanics, where they shed light on wavefunction dynamics, quantum
coherence, and the vibrational modes of particles in potential wells.
It is important to distinguish between these two concepts, which,
despite appearing similar, are fundamentally different. The Schrödinger
oscillator is derived by incorporating a non-minimal coupling of the
form $\overrightarrow{p}\rightarrow(\overrightarrow{p}-im\omega\overrightarrow{r})$
into the Schrödinger equation, $\overrightarrow{p}^{2}\Psi=i2m\partial_{t}\Psi$,
resulting in the Schrödinger oscillator \citep{key-3}. In contrast,
the harmonic oscillator describes a particle confined within a quadratic
(harmonic) potential. It is worth noting that the used momentum substitution
can be interpreted as a Weyl-type covariant derivative, analogous
to early formulations of gauge invariance. Consequently, the associated
non-Hermiticity is not arbitrary but possesses a well-defined geometrical
and gauge-related structure. This feature has been discussed in the
context of non-Hermitian quantum systems with generalized gauge symmetries
and provides an additional physical interpretation of the Schrödinger
oscillator formalism.

On the other side, in recent years, noncommutative (NC) geometry has
become an active theoretical area of research and has attracted considerable
interest in several domains of modern physics. Its influence extends
to quantum mechanics \citep{key-4}, quantum field theory \citep{key-5},
the Standard Model \citep{key-6}, string theory \citep{key-7},
matrix theory \citep{key-8}, as well as black hole physics, gravity,
and cosmology \citep{key-9,key-10}. NC spaces play a crucial role
in understanding physical phenomena at extremely short distances and
high-energy scales. Several types of NC spaces have been explored,
including time-dependent \citep{key-11}, position-dependent \citep{key-12,key-13},
energy-dependent \citep{key-14} and complex \citep{key-15} NC spaces.
However, of particular interest is the dynamical noncommutative (DNC)
space, also referred to as position-dependent NC space, in which the
NC parameter $\Theta$ is not constant but depends on the spatial
coordinates, $\Theta\rightarrow\Theta(X,Y)$. The transition from
the commutative framework to the NC one, or vice versa, can be achieved
through several approaches, including Weyl--Wigner map \citep{key-16},
Moyal--Weyl product ($\star$product) formalism \citep{key-17},
Seiberg--Witten map \citep{key-4}, and Bopp-shift transformation
\citep{key-18}. In this study, we rely on the Bopp-shift approach.

In this work, we explore the effect of DNC space on the Schrödinger
oscillator using perturbation theory and investigate its thermal properties.
Our goal is to deepen the understanding of NC spaces and their phenomenological
implications. This study builds upon our recent investigations within
the same framework. For instance, in \citep{key-19}, we analyzed
the two-dimensional (2D) Klein-Gordon oscillator in a DNC space, deriving
the energy eigenvalues and eigenvectors in both relativistic and nonrelativistic
regimes, while in \citep{key-20} we investigated its thermal properties.
Similarly, in \citep{key-21}, we examined the Landau problem in
a 2D DNC space and studied the magneto-conductivity using the Kubo
formula. In addition, in \citep{key-22}, we addressed the eigensystem
of the 2D DNC Dirac oscillator, whereas in \citep{key-23}, we studied
graphene within the DNC framework and explored its thermodynamic properties
in the extreme relativistic limit at zero temperature. From a physical
viewpoint, the investigation of thermal properties constitutes an
important step toward understanding the phenomenological consequences
of dynamical noncommutativity (NCy). Thermodynamic quantities such
as the partition function, internal energy, entropy, and specific
heat provide valuable information about the collective statistical
behavior of quantum systems and their response to deformation parameters.
In particular, studying these quantities for the Schrödinger oscillator
in a DNC framework enables us to examine how position-dependent NCy
modifies the thermal structure of the system and may lead to measurable
deviations from the standard commutative or constant-NC cases. Moreover,
such an analysis offers further insight into the interplay between
NC geometry, non-Hermitian structures, and quantum statistical physics.

Despite the considerable progress achieved in previous DNC studies
of relativistic and condensed-matter systems, including the Klein--Gordon
and Dirac oscillators, the Landau problem, and graphene systems, the
Schrödinger oscillator within this framework provides an opportunity
to further explore the phenomenological implications of dynamical
NCy in a nonrelativistic regime. In particular, this investigation
may reveal additional rich and nontrivial physical effects associated
with both dynamical and flat NC spaces beyond the relativistic setting.

The remainder of the paper is organized as follows. Sec. \ref{II}
reviews the DNC formalism. In Sec. \ref{III}, we investigate the
deformed Schrödinger oscillator; specifically, the Schrödinger oscillator
is extended to the DNC space in sub-Sec. \ref{III-A}. Subsequently,
in sub-Secs. \ref{III-B} and \ref{III.C}, we derive the eigensystem
of the Schrödinger oscillator in both dynamical and non-dynamical
NC spaces. Sec. \ref{IV} is devoted to thermal behavior analysis:
sub-Sec. \ref{IV-A} explores the thermal properties of the deformed
system using the partition function, while sub-Sec. \ref{IV-B} presents
the numerical results and corresponding discussion. Finally, Sec.
\ref{V} concludes the paper with remarks and future perspectives.

\section{\label{II}Fundamental aspects of DNC formalism}

At the extremely small scale known as the string scale, space becomes
NC, and the coordinate operators do not commute \citep{key-4}. Consequently,
the canonical variables obey a deformed Heisenberg commutation relation
of the form
\begin{equation}
\left[x_{\mu}^{nc},x_{\nu}^{nc}\right]=i\Theta_{\mu\nu},\label{eq:1}
\end{equation}
where $\Theta_{\mu\nu}$ is an anti-symmetric tensor. In the case
of a 2D NC system, it simplifies to $\Theta_{ij}=\Theta\epsilon_{ij}$
where $\epsilon_{ij}$ is the Levi-Civita symbol. In the simplest
formulation, the deformation parameter $\Theta$ is assumed to be
a constant real quantity. More generally, however, $\Theta$ can acquire
a functional dependence on the spatial coordinates \citep{key-23},
time \citep{key-11}, energy \citep{key-14}, or even hypermomentum
\citep{key-24}. In this work, we adopt a position-dependent NC space,
following the methodology developed in \citep{key-12,key-19}. Within
this framework, the position dependence of NCy is implemented through
a set of 2D non-Hermitian operators $X$, $Y$, $P_{X}$, and $P_{Y}$.
Accordingly, the NC parameter $\Theta$ is promoted to a coordinate-dependent
function, expressed as
\begin{equation}
\Theta\left(X,Y\right)=\Theta\left(1+\tau Y^{2}\right),\label{eq:2}
\end{equation}
where $\tau$ denotes the DNC parameter, often referred to as the
non-Hermitian parameter \citep{key-25}. The parameters $\Theta$
and $\tau$ have dimensions of $\text{L}^{2}$ and $\textrm{L}^{-2}$,
respectively. It is worth noting that alternative functional realizations
of $\Theta\left(X,Y\right)$ have appeared in prior studies. For instance,
Ref. \citep{key-26} considers the choice $\Theta\left(X,Y\right)=\Theta\left(1-\tau Y+\tau Y^{2}\right)$,
while in Ref. \citep{key-13} the deformation is given by $\Theta\left(X,Y\right)=\Theta\left[1+\Theta\alpha\left(1+X^{2}+Y^{2}\right)\right]$$^{-1}$,
where $\alpha$ represents a locality parameter.  A common feature
of these constructions is that the resulting coordinate operators
are inherently non-Hermitian. As discussed in \citep{key-27}, such
deformed NC geometries are closely connected to non-Hermitian Hamiltonian
frameworks  \citep{key-28}. In what follows, a consistent treatment
of the intrinsic non-Hermiticity is presented. In the DNC (or $\mathrm{\tau}$-deformed)
space, the non-Hermitian operators obey the commutation relations
\citep{key-12}:
\begin{equation}
\begin{array}{l}
\left[X,Y\right]=i\Theta\left(1+\tau Y^{2}\right),\;\left[X,P_{x}\right]=\left[Y,P_{y}\right]=i\hbar\left(1+\tau Y^{2}\right),\\
\left[X,P_{y}\right]=2i\tau Y\left(\Theta P_{y}+\hbar X\right),\;\left[Y,P_{x}\right]=\left[P_{x},P_{y}\right]=0.
\end{array}\label{eq:3}
\end{equation}

Moreover, in the DNC phase-space, the momentum sector exhibits a nontrivial
structure. In contrast to the vanishing commutator $\left[P_{x},P_{y}\right]$
appearing in Eq. (\ref{eq:3}), the momenta satisfy $\left[P_{x},P_{y}\right]=i\bar{\Theta}\left(1+\tau Y^{2}\right)$,
where $\bar{\Theta}$ denotes a constant NC parameter associated with
momentum and carries the dimension $\textrm{M}^{2}\textrm{L}^{2}\textrm{T}^{-1}.$
The algebra in Eq. (\ref{eq:3}) can be realized in in terms of the
standard Hermitian NC variable operators $x^{nc}$, $y^{nc}$, $p_{x}^{nc}$,
and $p_{y}^{nc}$ through the mapping:
\begin{equation}
X=\left(1+\tau\left(y^{nc}\right)^{2}\right)x^{nc},\;Y=y^{nc},\;P_{x}=p_{x}^{nc},\;P_{y}=\left(1+\tau\left(y^{nc}\right)^{2}\right)p_{y}^{nc}.\label{eq:4}
\end{equation}

In the limit $\tau\text{\textrightarrow}0$, and upon using Eq. (\ref{eq:4}),
Eq. (\ref{eq:3}) reduces to the standard NC (non-DNC) commutation
relations:
\begin{equation}
\begin{array}{l}
\left[x^{nc},y^{nc}\right]=i\Theta,\\
\left[x^{nc},p_{x}^{nc}\right]=\left[y^{nc},p_{y}^{nc}\right]=i\hbar,\\
\left[x^{nc},p_{y}^{nc}\right]=\left[y^{nc},p_{x}^{nc}\right]=\left[p_{x}^{nc},p_{y}^{nc}\right]=0.
\end{array}\label{eq:5}
\end{equation}

The representation (\ref{eq:4}) clearly indicates that some of the
involved operators are non-Hermitian and their adjoints satisfy
\begin{equation}
\begin{array}{cccc}
X^{\dagger}=X+2i\tau\Theta Y,\; & Y^{\dagger}=Y,\; & P_{x}^{\dagger}=P_{x}\;\textrm{and } & P_{y}^{\dagger}=P_{y}-2i\tau\hbar Y.\end{array}\label{eq:6}
\end{equation}

Consequently, a Hamiltonian constructed from these variables will,
in general, be non-Hermitian, i.e., $\mathscr{H}^{\dagger}\left(X_{i},P_{i}\right)\neq\mathscr{H}\left(X_{i},P_{i}\right)$.
To restore Hermiticity, we employ the Dyson map \citep{key-12} $\kappa=(1+\tau Y^{2})^{-\frac{1}{2}}$
which acts on any non-Hermitian observable $\mathcal{O}\neq\mathcal{O}^{\dagger}$
according to $\mathcal{O}^{\dagger}\overset{\kappa}{\rightarrow}$$\kappa\mathcal{O}\kappa^{-1}=o=o^{\dagger}$.
This transformation maps the non-Hermitian variables to an equivalent
set of Hermitian operators. It should be noted that the Dyson map
employed here is not unique. In general, different similarity transformations
may be used to relate a non-Hermitian Hamiltonian to an equivalent
Hermitian representation. However, the present choice is naturally
motivated by the specific form of the DNC algebra considered in Eq.
(\ref{eq:2}), particularly by the position-dependent deformation
factor $\left(1+\tau Y^{2}\right)$. Accordingly, for alternative
forms of the dynamical deformation, different Dyson maps may arise.
We also note that other frameworks for treating non-Hermitian systems
exist in the literature, including \ensuremath{\mathscr{P}}\ensuremath{\mathscr{T}}-symmetric
quantum mechanics, pseudo-Hermitian approaches, and formulations based
on biorthogonal inner products. While the explicit form of the observables
and the associated inner-product structure may depend on the chosen
framework, physically observable quantities such as the energy spectrum
are generally expected to remain invariant under equivalent similarity
transformations. The condition $\tau\geq0$ ensures that the Dyson
map remains well defined and nonsingular. Applying this procedure
yields a new set of Hermitian variables $x$, $y$, $p_{x}$, and$p_{y}$,
which can be expressed in terms of the NC variables as follows:
\begin{equation}
\begin{array}{l}
x=\kappa X\kappa^{-1}=(1+\tau\left(y^{nc}\right)^{2})^{\frac{1}{2}}x^{nc}(1+\tau\left(y^{nc}\right)^{2})^{\frac{1}{2}}=x^{\dagger},\\
y=\kappa Y\kappa^{-1}=y^{nc}=y^{\dagger},\\
p_{x}=\kappa P_{x}\kappa^{-1}=p_{x}^{nc}=p_{x}^{\dagger},\\
p_{y}=\kappa P_{y}\kappa^{-1}=(1+\tau\left(y^{nc}\right)^{2})^{\frac{1}{2}}p_{y}^{nc}(1+\tau\left(y^{nc}\right)^{2})^{\frac{1}{2}}=p_{y}^{\dagger}.
\end{array}\label{eq:7}
\end{equation}

The Hermitian DNC variables (\ref{eq:7}) satisfy the following deformed
commutation relations:
\begin{equation}
\begin{array}{l}
\left[x,y\right]=i\Theta\left(1+\tau y^{2}\right),\\
\left[x,p_{x}\right]=\left[y,p_{y}\right]=i\hbar\left(1+\tau Y^{2}\right),\\
\left[x,p_{y}\right]=2i\tau y\left(\Theta p_{y}+\hbar x\right),\\
\left[y,p_{x}\right]=0,\;\left[p_{x},p_{y}\right]=0.
\end{array}\label{eq:8}
\end{equation}

To establish a connection with ordinary quantum mechanics, the NC
variables may be mapped onto standard commutative variables via the
Bopp-shift \citep{key-4,key-18}:
\begin{equation}
\begin{array}{cccc}
x^{nc}=x^{s}-\frac{\Theta}{2\hbar}p_{y}^{s}\; & y^{nc}=y^{s}+\frac{\Theta}{2\hbar}p_{x}^{s},\; & p_{x}^{nc}=p_{x}^{s},\;\textrm{and } & p_{y}^{nc}=p_{y}^{s},\end{array}\label{eq:9}
\end{equation}
where the superscript $s$ denotes operators in the standard commutative
phase-space.

Within the DNC framework, for operators satisfying Eq. (\ref{eq:3}),
the generalized uncertainty principle takes the form $\triangle\mathscr{A}\triangle\mathscr{B}\geq\frac{1}{2}\left|\left\langle \left[\mathscr{A},\mathscr{B}\right]\right\rangle _{\rho}\right|$,
with $\mathscr{A},\mathscr{B}$ $\in\left\{ X,Y,P_{X},P_{Y}\right\} $.
As a consequence, a nonvanishing minimal uncertainty arises for the
coordinate $X$ in a simultaneous measurement of $X$ and $Y$ \citep{key-12},
\begin{equation}
\triangle X_{\textrm{min}}=\Theta\sqrt{\tau}\sqrt{1+\tau\left\langle Y\right\rangle _{\rho}^{2}},\label{eq:10}
\end{equation}
whereas no minimal length is associated with the coordinate $Y$.
Here, $\rho=\kappa^{2}=1/(1+\tau Y^{2})$ denotes the metric operator,
which is Hermitian, and $\left\langle Y\right\rangle _{\rho}$ represents
the expectation value of $Y$ with respect to this metric,
\begin{equation}
\left\langle Y\right\rangle _{\rho}=\left\langle \Psi\right|Y\left|\Psi\right\rangle _{\rho}.\label{eq:11}
\end{equation}

The corresponding inner product is defined by 
\begin{equation}
\left\langle \Phi\mid\Psi\right\rangle _{\rho}:=\left\langle \Phi\mid\rho\Psi\right\rangle =\int_{-\infty}^{+\infty}\frac{dy}{1+\tau y^{2}}\Psi^{*}\left(y\right)\Phi\left(y\right),\label{eq:12}
\end{equation}
for arbitrary states $\left|\Phi\right\rangle $ and $\left|\Psi\right\rangle $.
Any observable $\mathscr{P}$ must therefore satisfy $\left\langle \Phi\mid\mathscr{P}\Psi\right\rangle _{\rho}=\left\langle \mathscr{P}\Phi\mid\Psi\right\rangle _{\rho}$,
ensuring its Hermiticity with respect to the metric $\rho$. For the
non-Hermitian position operators $X$ and  $Y,$ the uncertainty relation
implied by Eq. (\ref{eq:12}), becomes $\triangle X\triangle Y\geq\frac{1}{2}\left|i\Theta\left\langle 1+\tau Y^{2}\right\rangle \right|$.
Furthermore, a minimal uncertainty in momentum emerges in a simultaneous
measurement of $Y$ and $P_{y}$ \citep{key-12}:
\begin{equation}
\triangle\left(P_{y}\right)_{\textrm{min}}=\hbar\sqrt{\tau}\sqrt{1+\tau\left\langle Y\right\rangle _{\rho}^{2}},\label{eq:13}
\end{equation}
while no minimal length or minimal momentum appears in a simultaneous
measurements of $X$ and $P_{x}$. The solutions corresponding to
simultaneous measurements of $Y$ and $P_{y}$ are given by
\begin{equation}
\triangle Y=\frac{1}{\hbar\tau}\triangle P_{y}\pm\frac{1}{\hbar\tau}\sqrt{\triangle P_{y}^{2}-\hbar^{2}\tau\left(1+\tau\left\langle Y\right\rangle ^{2}\right)}.\label{eq:14}
\end{equation}

A significant physical implication of position-dependent NCy is the
emergence of string-like behavior for objects in 2D spaces \citep{key-12}.
In particular, the existence of a minimal length for the coordinate
$X$ in simultaneous $X\text{\textendash}Y$ measurements implies
a fundamental limit to spatial resolution. This suggests that objects
defined in the $(X,Y)$ plane may possess an intrinsic extended, string-like
character. Consequently, DNC spaces exhibit a deeper and more natural
connection to string theory than their constant-NC counterparts. Moreover,
investigating the dynamics of systems such as the Schrödinger oscillator
within the DNC framework provides insight into the interplay between
position-dependent NCy and non-Hermitian quantum systems, while also
offering a useful setting for exploring possible effective consequences
of string-inspired geometrical structures. Such studies may provide
further theoretical insight into the role of position-dependent NCy
and its possible phenomenological implications beyond the standard
NC framework.   

It is important to emphasize that, although DNC geometry is motivated
by concepts originating from string theory and quantum gravity, the
present work is not intended to describe physics directly at the fundamental
string scale. Rather, the Schrödinger oscillator is employed here
as an effective nonrelativistic quantum model that provides a mathematically
tractable framework for investigating the physical consequences of
position-dependent NCy. In this sense, the string-like behavior emerging
in DNC spaces should be understood as a geometric feature associated
with the underlying NC algebra and the existence of a minimal length
scale, rather than as evidence that the oscillator itself represents
a fundamental string excitation. Consequently, the use of the nonrelativistic
Schrödinger formalism is appropriate for exploring low-energy and
effective manifestations of DNC geometry. Moreover, the mass parameter
$m$ appearing in the Hamiltonian is interpreted as the effective
particle mass within the oscillator model and is not assumed to be
related to the string or Planck mass scales.

\section{{\normalsize\label{III}2D Schrödinger oscillator in DNC space}}

\subsection{{\normalsize\label{III-A}}Extension to DNC space}

The Schrödinger oscillator in commutative space is described as follows
\citep{key-3}:
\begin{equation}
\frac{1}{2m}\underset{\overrightarrow{\Pi}^{\dagger}}{\underbrace{\left(\overrightarrow{p}^{s}+im\omega\overrightarrow{r}^{s}\right)}}\cdot\underset{\overrightarrow{\Pi}}{\underbrace{\left(\overrightarrow{p}^{s}-im\omega\overrightarrow{r}^{s}\right)}}\psi\left(\overrightarrow{r}^{s}\right)=E\psi\left(\overrightarrow{r}^{s}\right),\label{eq:15}
\end{equation}
where $\omega$ is the oscillator angular frequency, $m$ is the particle's
mass, $\overrightarrow{p}^{s}=-i\hbar\overrightarrow{\nabla}$. Note
that an electromagnetic interaction can be incorporated into Eq. (\ref{eq:15})
by making the usual minimal substitution $\overrightarrow{p}^{s}\rightarrow\left(\overrightarrow{p}^{s}-e\overrightarrow{A}_{s}\right)$,
where $\overrightarrow{A}_{s}$ is the electromagnetic potential. 

We extend the calculation to two dimensions, resulting in the following
equation:
\begin{equation}
\mathscr{H}\left(x_{i}^{s},p_{i}^{s}\right)\psi\left(\overrightarrow{r}^{s}\right)=E\psi\left(\overrightarrow{r}^{s}\right),\label{eq:16}
\end{equation}
where the scalar radial coordinate is defined as  $r^{s}=\left|\overrightarrow{r}^{s}\right|=\sqrt{\left(x^{s}\right)^{2}+\left(y^{s}\right)^{2}}$,
and the Schrödinger oscillator Hamiltonian is
\begin{equation}
\mathscr{H}\left(x_{i}^{s},p_{i}^{s}\right)=\frac{1}{2m}\left\{ \left(p_{x}^{s}+im\omega x^{s}\right)\left(p_{x}^{s}-im\omega x^{s}\right)+\left(p_{y}^{s}+im\omega y^{s}\right)\left(p_{y}^{s}-im\omega y^{s}\right)\right\} .\label{eq:17}
\end{equation}

Through direct calculations, we obtain
\begin{equation}
\mathscr{H}\left(x_{i}^{s},p_{i}^{s}\right)=\frac{1}{2m}\left\{ \left(p_{x}^{s}\right)^{2}+\left(p_{y}^{s}\right)^{2}+m^{2}\omega^{2}\left(\left(x^{s}\right)^{2}+\left(y^{s}\right)^{2}\right)-2m\omega\hbar\right\} .\label{eq:18}
\end{equation}

Therefore, in the DNC space, the Schrödinger oscillator Hamiltonian
transforms into
\begin{equation}
\mathscr{H}\left(x_{i},p_{i}\right)=\frac{1}{2m}\left\{ p_{x}^{2}+p_{y}^{2}+m^{2}\omega^{2}\left(x^{2}+y^{2}\right)-2m\omega\hbar\right\} .\label{eq:19}
\end{equation}

Now, by utilizing Eq. (\ref{eq:7}), we rewrite the Schrödinger oscillator
Hamiltonian (\ref{eq:19}) in terms of the NC variables
\begin{equation}
\begin{array}{l}
\mathscr{H}\left(x_{i}^{nc},p_{i}^{nc}\right)=\frac{1}{2m}\left\{ \left(p_{x}^{nc}\right)^{2}+\left(1+\tau\left(y^{nc}\right)^{2}\right)^{\frac{1}{2}}p_{y}^{nc}\left(1+\tau\left(y^{nc}\right)^{2}\right)p_{y}^{nc}\left(1+\tau\left(y^{nc}\right)^{2}\right)^{\frac{1}{2}}\right.\\
+\left.m^{2}\omega^{2}\left(\left(1+\tau\left(y^{nc}\right)^{2}\right)^{\frac{1}{2}}x^{nc}\left(1+\tau\left(y^{nc}\right)^{2}\right)x^{nc}\left(1+\tau\left(y^{nc}\right)^{2}\right)^{\frac{1}{2}}+\left(y^{nc}\right)^{2}\right)-2m\omega\hbar\right\} .
\end{array}\label{eq:20}
\end{equation}

Since $\tau$ is sufficiently small, the expression within parentheses
in Eq. (\ref{eq:20}) can be expanded to first-order as follows:
\begin{equation}
\left(1+\tau\left(y^{nc}\right)^{2}\right)^{\frac{1}{2}}=1+\frac{1}{2}\tau\left(y^{nc}\right)^{2}+\mathcal{O}\left(\tau^{2}\right),\label{eq:21}
\end{equation}
subsequently, Eq. (\ref{eq:20}) takes the form
\begin{equation}
\begin{array}{l}
\mathscr{H}\left(x_{i}^{nc},p_{i}^{nc}\right)=\frac{1}{2m}\left\{ \left(p_{x}^{nc}\right)^{2}+\left(p_{y}^{nc}\right)^{2}+\tau p_{y}^{nc}\left(y^{nc}\right)^{2}p_{y}^{nc}+\frac{\tau}{2}\left(y^{nc}\right)^{2}\left(p_{y}^{nc}\right)^{2}+\frac{\tau}{2}\left(p_{y}^{nc}\right)^{2}\left(y^{nc}\right)^{2}\right.\\
+\left.m^{2}\omega^{2}\left\{ \left(x^{nc}\right)^{2}+\left(y^{nc}\right)^{2}+\tau x^{nc}\left(y^{nc}\right)^{2}x^{nc}+\frac{\tau}{2}\left(y^{nc}\right)^{2}\left(x^{nc}\right)^{2}+\frac{\tau}{2}\left(x^{nc}\right)^{2}\left(y^{nc}\right)^{2}\right\} -2m\omega\hbar\right\} .
\end{array}\label{eq:22}
\end{equation}

Using the Bopp-shift transformation (\ref{eq:9}), the Hamiltonian
(\ref{eq:22}) can now be written in terms of the standard commutative
variables as follows:
\begin{equation}
\begin{array}{l}
\mathscr{H}\left(x_{i}^{s},p_{i}^{s}\right)=\frac{1}{2m}\left\{ \left(p_{x}^{s}\right)^{2}+\left(p_{y}^{s}\right)^{2}+\tau p_{y}^{s}\left(y^{s}+\frac{\Theta}{2\hbar}p_{x}^{s}\right)^{2}p_{y}^{s}+\frac{\tau}{2}\left(y^{s}+\frac{\Theta}{2\hbar}p_{x}^{s}\right)^{2}\left(p_{y}^{s}\right)^{2}\right.\\
+\left.\frac{\tau}{2}\left(p_{y}^{s}\right)^{2}\left(y^{s}+\frac{\Theta}{2\hbar}p_{x}^{s}\right)^{2}\right\} +m^{2}\omega^{2}\left\{ \left(x^{s}-\frac{\Theta}{2\hbar}p_{y}^{s}\right)^{2}+\left(y^{s}+\frac{\Theta}{2\hbar}p_{x}^{s}\right)^{2}+\frac{\tau}{2}\left(y^{s}+\frac{\Theta}{2\hbar}p_{x}^{s}\right)^{2}\left(x^{s}-\frac{\Theta}{2\hbar}p_{y}^{s}\right)^{2}\right.\\
+\left.\frac{\tau}{2}\left(x^{s}-\frac{\Theta}{2\hbar}p_{y}^{s}\right)^{2}\left(y^{s}+\frac{\Theta}{2\hbar}p_{x}^{s}\right)^{2}+\tau\left(x^{s}-\frac{\Theta}{2\hbar}p_{y}^{s}\right)\left(y^{s}+\frac{\Theta}{2\hbar}p_{x}^{s}\right)^{2}\left(x^{s}-\frac{\Theta}{2\hbar}p_{y}^{s}\right)\right\} -2m\omega\hbar+\mathcal{O}\left(\tau^{2}\right).
\end{array}\label{eq:23}
\end{equation}

It is important to note that, to first order in $\Theta$ and $\tau$,
Eq. (\ref{eq:23}) simplifies to
\begin{equation}
\begin{array}{l}
\mathscr{H}\left(x_{i}^{s},p_{i}^{s}\right)=\frac{1}{2m}\left\{ \left(p_{x}^{s}\right)^{2}+\left(p_{y}^{s}\right)^{2}+m^{2}\omega^{2}\left\{ \left(x^{s}\right)^{2}+\left(y^{s}\right)^{2}\right\} -2m\hbar\omega\right\} \\
+\frac{\tau}{2m}\left\{ \frac{1}{2}\left(p_{y}^{s}\right)^{2}\left(y^{s}\right)^{2}+p_{y}^{s}\left(y^{s}\right)^{2}p_{y}^{s}+\frac{1}{2}\left(y^{s}\right)^{2}\left(p_{y}^{s}\right)^{2}+\frac{1}{2}m^{2}\omega^{2}\left(y^{s}\right)^{2}\left(x^{s}\right)^{2}\right.\\
+\left.\frac{1}{2}m^{2}\omega^{2}\left(x^{s}\right)^{2}\left(y^{s}\right)^{2}+m^{2}\omega^{2}\left(x^{s}\right)^{2}\left(y^{s}\right)^{2}\right\} -\Theta\frac{1}{2\hbar}m\omega^{2}L_{z}+\mathcal{O}\left(\Theta^{2}\right),
\end{array}\label{eq:24}
\end{equation}
where terms involving $\Theta\tau$ are also neglected. Eq. (\ref{eq:24})
can now be expressed as:
\begin{equation}
\mathscr{H}=\mathscr{H}^{(0)}+\mathscr{H}^{(\Theta)}+\mathscr{H}^{(\tau)}+\mathcal{O}\left(\Theta\tau,\Theta^{2},\tau^{2}\right),\label{eq:25}
\end{equation}
with
\begin{equation}
\mathscr{H}^{(0)}=\frac{1}{2m}\left\{ \left(p_{x}^{s}\right)^{2}+\left(p_{y}^{s}\right)^{2}\right\} +\frac{1}{2}m\omega^{2}\left\{ \left(x^{s}\right)^{2}+\left(y^{s}\right)^{2}\right\} -\hbar\omega,\label{eq:26}
\end{equation}
\begin{equation}
\mathscr{H}^{(\Theta)}=-\frac{\Theta}{2\hbar}m\omega^{2}L_{z},\label{eq:27}
\end{equation}
\begin{equation}
\mathscr{H}^{(\tau)}=\frac{\tau}{2m}\left\{ \frac{1}{2}\left(p_{y}^{s}\right)^{2}\left(y^{s}\right)^{2}+\frac{1}{2}\left(y^{s}\right)^{2}\left(p_{y}^{s}\right)^{2}+p_{y}^{s}\left(y^{s}\right)^{2}p_{y}^{s}+2m^{2}\omega^{2}\left(x^{s}\right)^{2}\left(y^{s}\right)^{2}\right\} ,\label{eq:28}
\end{equation}
where
\begin{equation}
L_{z}=p_{y}^{s}x^{s}-p_{x}^{s}y^{s}=\left(\overrightarrow{r}^{s}\times\overrightarrow{p}^{s}\right)_{z}.\label{eq:29}
\end{equation}

In the following steps, we will solve the energy equation given by
\begin{equation}
\mathscr{H}\left|\psi\right\rangle =\left\{ \mathscr{H}^{(0)}+\mathscr{H}^{(\Theta)}+\mathscr{H}^{(\tau)}\right\} \left|\psi\right\rangle =E^{\textrm{DNC}}\left|\psi\right\rangle ,\label{eq:30}
\end{equation}
where the wave function of the system is given by
\begin{equation}
\left|\psi\right\rangle ^{\textrm{t}}=\left(\begin{array}{cc}
\left|\psi_{1}\right\rangle  & \left|\psi_{2}\right\rangle \end{array}\right),\label{eq:31}
\end{equation}
where $\textrm{t}$ denotes the transpose. Next, we explore the influence
of the perturbed Hamiltonians on the Schrödinger oscillator. Since
the DNC and NC parameters $\tau$ and $\Theta$ are both non-zero
and sufficiently small, perturbation theory can be applied to determine
the system's eigenvalues and eigenstates.

\subsection{{\normalsize\label{III-B}}The Schrödinger oscillator in NC space:
unperturbed system}

The Schrödinger oscillator in NC space is described by the equation
\begin{equation}
\left\{ \mathscr{H}^{(0)}+\mathscr{H}^{(\Theta)}\right\} \left|\psi\right\rangle =E^{\textrm{NC}}\left|\psi\right\rangle ,\label{eq:32}
\end{equation}
where $E^{\textrm{NC}}$ represents the eigenenergy of the NC Schrödinger
oscillator. By employing Eqs. (\ref{eq:26}, \ref{eq:27}), we obtain
\begin{equation}
\left\{ \frac{1}{2m}\left\{ \left(p_{x}^{s}\right)^{2}+\left(p_{y}^{s}\right)^{2}\right\} +\frac{1}{2}m\omega^{2}\left\{ \left(x^{s}\right)^{2}+\left(y^{s}\right)^{2}\right\} -\hbar\omega-\Theta\frac{1}{2\hbar}m\omega^{2}L_{z}\right\} \left|\psi\right\rangle =E^{\textrm{NC}}\left|\psi\right\rangle .\label{eq:33}
\end{equation}

We conveniently introduce the ladder operators $a_{\pm}$ as follows:
\begin{equation}
a_{\pm}=\frac{1}{2}\left\{ \sqrt{\frac{m\omega}{\hbar}}\left(x^{s}\pm iy^{s}\right)+i\sqrt{\frac{1}{m\omega\hbar}}\left(p_{x}^{s}\pm ip_{y}^{s}\right)\right\} ,\label{eq:34}
\end{equation}
with the commutation relation
\begin{equation}
\left[a_{\pm},a_{\pm}^{\dagger}\right]=1,\label{eq:35}
\end{equation}
and the number operators 
\begin{equation}
N_{\pm}=a_{\pm}^{\dagger}a_{\pm},\textrm{ with }\:N=N_{+}+N_{-}.\label{eq:36}
\end{equation}

We then express $\mathscr{H}^{(0)}+\mathscr{H}^{(\Theta)}$ in terms
of Eq. (\ref{eq:36}) as follows:
\begin{equation}
\mathscr{H}^{(0)}+\mathscr{H}^{(\Theta)}=\hbar\omega\left(N+1\right)-\frac{1}{2}m\omega^{2}\Theta\left(N_{-}-N_{+}\right),\label{eq:37}
\end{equation}
where
\begin{equation}
L_{z}=\hbar\left(N_{-}-N_{+}\right).\label{eq:38}
\end{equation}

It should be noted that the common eigenvectors of $\mathscr{H}$
and $L_{z}$ are $\mid n_{+},n_{-}>$, with
\begin{equation}
N\mid n_{+},n_{-}>=n\mid n_{+},n_{-}>,\label{eq:39}
\end{equation}
and
\begin{equation}
L_{z}\mid n_{+},n_{-}>=l\mid n_{+},n_{-}>,\label{eq:40}
\end{equation}
where
\begin{equation}
\left|\psi\right\rangle \equiv\mid n_{+},n_{-}>=\frac{\left(a_{+}^{\dagger}\right)^{n_{+}}\left(a_{-}^{\dagger}\right)^{n_{-}}}{\sqrt{\left(n_{+}!\right)\left(n_{-}!\right)}}\mid0,0>,\label{eq:41}
\end{equation}
with $\mid0,0>$ is the fundamental state. Note that the eigenfunctions
(\ref{eq:41}) are \ensuremath{\mathscr{P}}\ensuremath{\mathscr{T}}-symmetric
ensuring no broken \ensuremath{\mathscr{P}}\ensuremath{\mathscr{T}}-regime.
The effects of $\mathscr{H}^{(\Theta)}$ on the Schrödinger oscillator
system through the energy eigenvalues are given as follows:
\begin{equation}
E_{n,l}^{\textrm{NC}}=\left(n+1\right)\hbar\omega-\Theta\frac{1}{2}m\omega^{2}l-\hbar\omega.\label{eq:42}
\end{equation}
It should be noted that the first term in the energy eigenvalues (\ref{eq:42})
resembles that of the harmonic oscillator. Furthermore, it is evident
that the presence of the parameter $\Theta$ in Eq. (\ref{eq:42})
breaks the degeneracy of the energy levels.  In the limit as $\Theta\rightarrow0,$
the system returns to the unperturbed case, and the energy eigenvalues
are given by
\begin{equation}
E_{n}=\hbar\omega\left(n+1\right)-\hbar\omega.\label{eq:43}
\end{equation}

\begin{figure}[H]
\centering{}%
\begin{tabular}{cc}
\includegraphics[scale=0.73]{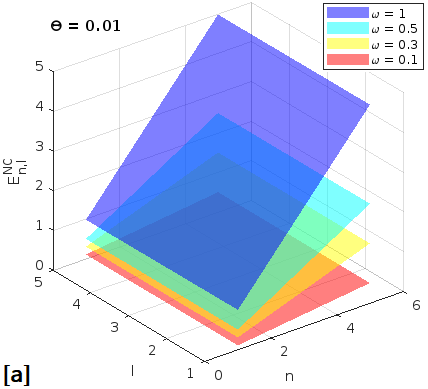} & \includegraphics[scale=0.74]{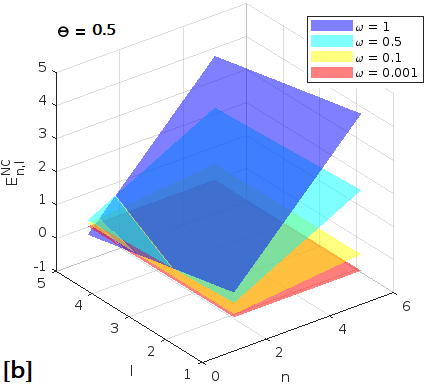}\tabularnewline
\end{tabular}\caption{\label{1}3D plots of the NC energy eigenvalues $E_{n,l}^{\textrm{NC}}$
as a function of quantum numbers $n$ \& $l$, shown for different
values of $\omega$, with $\Theta$ fixed. (a) $\Theta=0.01$ ; (b)
$\Theta=0.5$. }
\end{figure}

\begin{figure}[H]
\centering{}%
\begin{tabular}{cc}
\includegraphics[scale=0.58]{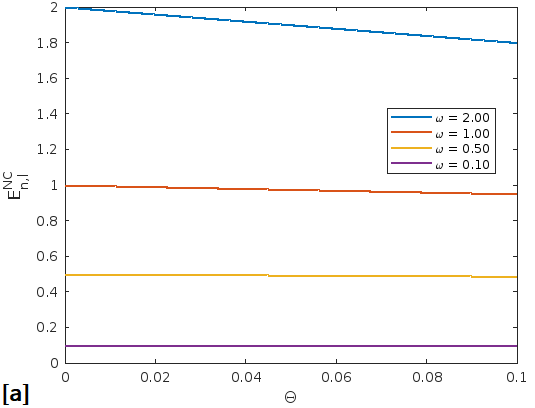} & \includegraphics[scale=0.58]{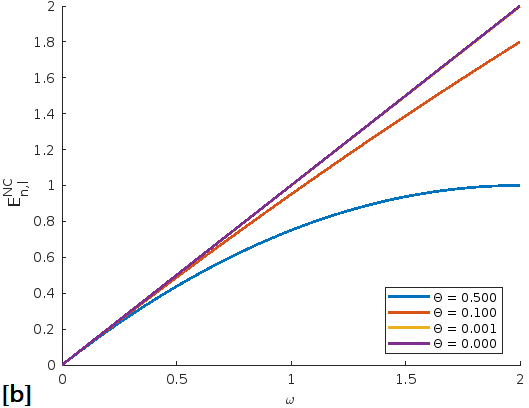}\tabularnewline
\end{tabular}\caption{\label{2} (a) Energy eigenvalues $E_{n,l}^{\textrm{NC}}$ as a function
of $\Theta,$ shown for different values of the oscillator angular
frequency $\omega$. (b) $E_{n,l}^{\textrm{NC}}$ as a function of
$\omega$, shown for different values of $\Theta.$}
\end{figure}

In Fig. \ref{1}, we depict the spectrum of the system as a function
of the quantum numbers $n$ \& $l$ for two different values of $\Theta$.
Sub-Fig. (\ref{2}.a) shows the eigenenergy vs $\Theta$ for different
values of $\omega$. In sub-Fig (\ref{2}.b), we plot the eigenenergy
vs $\omega$ for different values of $\Theta$. Note that $\hbar=m=1$.

From the plots in Fig. (\ref{2}), we observe a significant modification
of the energy spectrum due to $\Theta$. Small values of $\Theta$
yield near-classical behavior, aligning with the commutative case.
In contrast, larger $\Theta$ introduces sharp deviations, elevating
energy levels and amplifying the NC geometry role. Higher $\omega$
further enhance these NC effects, steepening energy scaling with both
$\Theta$ and $n$, $l$. These shifts highlight the nontrivial role
of NC deformation in reshaping the quantum energy landscape, with
measurable implications for systems where tuning $\omega$ or $\Theta$
could probe space NCy. 

\subsection{{\normalsize\label{III.C}}Perturbed system}

Since the DNC parameter $\tau$ is significantly smaller than the
system's energy scales, we consider its effect as a perturbation.
To analyze the system in $\tau$-deformed space, we apply the time-independent
perturbation theory. If $\tau$ is nonzero, it must remain very small. 

From Eq. (\ref{eq:34}), we obtain
\begin{equation}
\begin{cases}
x^{s}=\frac{1}{2}\sqrt{\frac{\hbar}{m\omega}}\left(a_{+}+a_{+}^{\dagger}+a_{-}+a_{-}^{\dagger}\right),\\
y^{s}=\frac{1}{2i}\sqrt{\frac{\hbar}{m\omega}}\left(a_{+}-a_{+}^{\dagger}-a_{-}+a_{-}^{\dagger}\right),
\end{cases}\:\begin{cases}
p_{x}^{s}=\frac{i}{2}\sqrt{m\omega\hbar}\left(-a_{+}+a_{+}^{\dagger}-a_{-}+a_{-}^{\dagger}\right),\\
p_{y}^{s}=\frac{1}{2}\sqrt{m\omega\hbar}\left(-a_{+}-a_{+}^{\dagger}+a_{-}+a_{-}^{\dagger}\right).
\end{cases}\label{eq:44}
\end{equation}

By defining $n=n_{+}+n_{-}$ and $l=n_{+}-n_{-}$, the eigenkets of
the Hamiltonian can be expressed as
\begin{equation}
\mid n_{+},n_{-}>=\mid\frac{n+l}{2},\frac{n-l}{2}>,\label{eq:45}
\end{equation}
enabling the classification of the following states:
\begin{center}
\begin{tabular}{lll}
\toprule 
 & $n$, $l$ & Eigenkets \tabularnewline
\midrule
\midrule 
Ground state & $n=0$, $l=0$ & $\mathrm{\mid0,0>}$\tabularnewline
\midrule 
\multirow{2}{*}{1st excited state} & $n=1$, $l=1$ & $\mathrm{\mid1,0>}$\tabularnewline
\cmidrule{2-3}
 & $n=1$, $l=-1$ & $\mathrm{\mid0,1>}$\tabularnewline
\midrule 
\multirow{3}{*}{2nd excited state} & $n=2$, $l=2$ & $\mid2,0>$\tabularnewline
\cmidrule{2-3}
 & $n=2$, $l=0$ & $\mathrm{\mid1,1>}$\tabularnewline
\cmidrule{2-3}
 & $n=2,l=-2$ & $\mid0,2>$\tabularnewline
\bottomrule
\end{tabular}
\par\end{center}

$\phantom{}$

We primarily focus on determining $E_{n}^{(1)}$ , which corresponds
to the first-order corrections to the unperturbed system. However,
we begin by analyzing the impact of the perturbed Hamiltonian on the
energy of the ground state, i.e., $n=0$, $l=0$, specifically by
evaluating $\left\langle 0,0\right|\mathscr{H}^{(\tau)}\left|0,0\right\rangle $.
Now, let us compute the individual terms in Eq. (\ref{eq:28}), starting
with the first term, $\mathscr{T}_{1}=\frac{\tau}{4m}\left(p_{y}^{s}\right)^{2}\left(y^{s}\right)^{2}$.
By applying Eq. (\ref{eq:44}), we derive the following:
\begin{equation}
\begin{array}{l}
\left(p_{y}^{s}\right)^{2}=\frac{1}{4}m\omega\hbar\left\{ a_{+}^{2}+a_{+}a_{+}^{\dagger}-a_{+}a_{-}-a_{+}a_{-}^{\dagger}+a_{+}^{\dagger}a_{+}+\left(a_{+}^{\dagger}\right)^{2}-a_{+}^{\dagger}a_{-}\right.\\
\left.-a_{+}^{\dagger}a_{-}^{\dagger}-a_{-}a_{+}-a_{-}a_{+}^{\dagger}+a_{-}^{2}+a_{-}a_{-}^{\dagger}+-a_{-}^{\dagger}a_{+}-a_{-}^{\dagger}a_{+}^{\dagger}+a_{-}^{\dagger}a_{-}+\left(a_{-}^{\dagger}\right)^{2}\right\} ,
\end{array}\label{eq:46}
\end{equation}
and
\begin{equation}
\begin{array}{l}
\left(y^{s}\right)^{2}=-\frac{1}{4}\frac{\hbar}{m\omega}\left\{ a_{+}^{2}-a_{+}a_{+}^{\dagger}-a_{+}a_{-}+a_{+}a_{-}^{\dagger}-a_{+}^{\dagger}a_{+}+\left(a_{+}^{\dagger}\right)^{2}+a_{+}^{\dagger}a_{-}\right.\\
\left.-a_{+}^{\dagger}a_{-}^{\dagger}-a_{-}a_{+}+a_{-}a_{+}^{\dagger}+a_{-}^{2}-a_{-}a_{-}^{\dagger}+a_{-}^{\dagger}a_{+}-a_{-}^{\dagger}a_{+}^{\dagger}-a_{-}^{\dagger}a_{-}+\left(a_{-}^{\dagger}\right)^{2}\right\} .
\end{array}\label{eq:47}
\end{equation}

Subsequently,
\begin{equation}
\begin{array}{l}
\mathscr{T}_{1}=-\frac{1}{64}\frac{\hbar^{2}}{m}\tau\left\{ a_{+}^{2}\left(a_{+}^{\dagger}\right)^{2}-a_{+}a_{+}^{\dagger}a_{+}a_{+}^{\dagger}-a_{+}a_{+}^{\dagger}a_{-}a_{-}^{\dagger}+a_{+}a_{-}a_{+}^{\dagger}a_{-}^{\dagger}\right.\\
\left.+a_{+}a_{-}a_{-}^{\dagger}a_{+}^{\dagger}+a_{-}a_{+}a_{+}^{\dagger}a_{-}^{\dagger}+a_{-}a_{+}a_{-}^{\dagger}a_{+}^{\dagger}-a_{-}a_{-}^{\dagger}a_{+}a_{+}^{\dagger}-a_{-}a_{-}^{\dagger}a_{-}a_{-}^{\dagger}+a_{-}^{2}\left(a_{-}^{\dagger}\right)^{2}\right\} .
\end{array}\label{eq:48}
\end{equation}

Clearly, $\mathscr{T}_{1}$ consists of 256 terms, but only the selected
contributions above are nonzero. Next, we determine the correction
induced by this term on the ground-state energy of the system by evaluating
$\left\langle 0,0\right|\mathscr{T}_{1}\left|0,0\right\rangle $.
Through detailed calculations, we obtain $\left\langle 0,0\right|a_{+}^{2}\left(a_{+}^{\dagger}\right)^{2}\left|0,0\right\rangle =2$,
$\left\langle 0,0\right|a_{+}a_{+}^{\dagger}a_{+}a_{+}^{\dagger}\left|0,0\right\rangle =1$,
$\left\langle 0,0\right|a_{-}a_{-}^{\dagger}a_{-}a_{-}^{\dagger}\left|0,0\right\rangle =1$,
$\left\langle 0,0\right|a_{+}a_{-}a_{+}^{\dagger}a_{-}^{\dagger}\left|0,0\right\rangle =1$,
$\left\langle 0,0\right|a_{-}^{2}\left(a_{-}^{\dagger}\right)^{2}\left|0,0\right\rangle =2$,
leading to
\begin{equation}
\left\langle 0,0\right|\mathscr{T}_{1}\left|0,0\right\rangle =-\frac{1}{16}\frac{\hbar^{2}}{m}\tau.\label{eq:49}
\end{equation}

Using the same approach, we can determine the contributions of the
remaining terms to the ground-state energy of the system. Now, let
us examine the term $\mathscr{T}_{2}=\frac{\tau}{4m}\left(y^{s}\right)^{2}\left(p_{y}^{s}\right)^{2}$.
As before, this expression consists of 256 terms, but only the following
contributions are nonzero:
\begin{equation}
\begin{array}{l}
\mathscr{T}_{2}=-\frac{\tau}{4m}\frac{\hbar^{2}}{16}\left\{ a_{+}^{2}\left(a_{+}^{\dagger}\right)^{2}-a_{+}a_{+}^{\dagger}a_{+}a_{+}^{\dagger}-a_{+}a_{+}^{\dagger}a_{-}a_{-}^{\dagger}+a_{+}a_{-}a_{+}^{\dagger}a_{-}^{\dagger}\right.\\
\left.+a_{-}a_{+}a_{+}^{\dagger}a_{-}^{\dagger}+a_{-}a_{+}a_{-}^{\dagger}a_{+}^{\dagger}-a_{-}a_{-}^{\dagger}a_{+}a_{+}^{\dagger}-a_{-}a_{-}^{\dagger}a_{-}a_{-}^{\dagger}+a_{+}a_{-}a_{-}^{\dagger}a_{+}^{\dagger}+a_{-}^{2}\left(a_{-}^{\dagger}\right)^{2}\right\} .
\end{array}\label{eq:50}
\end{equation}

Then, we obtain
\begin{equation}
\left\langle 0,0\right|\mathscr{T}_{2}\left|0,0\right\rangle =\left\langle 0,0\right|-\frac{\hbar^{2}}{16m}\tau\left|0,0\right\rangle =-\frac{1}{16}\frac{\hbar^{2}}{m}\tau.\label{eq:51}
\end{equation}

Next, we examine the term $\mathscr{T}_{3}=\frac{\tau}{2m}p_{y}^{s}\left(y^{s}\right)^{2}p_{y}^{s}$,
where the non-zero contributions are given by
\begin{equation}
\begin{array}{l}
\mathscr{T}_{3}=-\frac{1}{32}\frac{\hbar^{2}}{m}\tau\left\{ -a_{+}a_{+}^{\dagger}a_{+}a_{+}^{\dagger}-a_{+}a_{-}a_{-}^{\dagger}a_{+}^{\dagger}-a_{-}a_{+}a_{-}^{\dagger}a_{+}^{\dagger}-a_{-}a_{-}^{\dagger}a_{+}a_{+}^{\dagger}\right.\\
\left.-a_{+}^{2}\left(a_{+}^{\dagger}\right)^{2}-a_{+}a_{+}^{\dagger}a_{-}a_{-}^{\dagger}-a_{+}a_{-}a_{+}^{\dagger}a_{-}^{\dagger}-a_{-}a_{+}a_{+}^{\dagger}a_{-}^{\dagger}-a_{-}^{2}\left(a_{-}^{\dagger}\right)^{2}-a_{-}a_{-}^{\dagger}a_{-}a_{-}^{\dagger}\right\} ,
\end{array}\label{eq:52}
\end{equation}
consequently, we obtain
\begin{equation}
\left\langle 0,0\right|\mathscr{T}_{3}\left|0,0\right\rangle =\frac{3}{8}\frac{\hbar^{2}}{m}\tau.\label{eq:53}
\end{equation}

Also, we examine $\mathscr{T}_{4}=\tau m\omega^{2}\left(x^{s}\right)^{2}\left(y^{s}\right)^{2}$,
where the non-zero contributions are given by
\begin{equation}
\begin{array}{l}
\mathscr{T}_{4}=\frac{-1}{16}\frac{\hbar^{2}}{m}\tau\left\{ a_{+}^{2}\left(a_{+}^{\dagger}\right)^{2}-a_{+}a_{+}^{\dagger}a_{+}a_{+}^{\dagger}-a_{+}a_{+}^{\dagger}a_{-}a_{-}^{\dagger}-a_{+}a_{-}a_{+}^{\dagger}a_{-}^{\dagger}\right.\\
-\left.a_{+}a_{-}a_{-}^{\dagger}a_{+}^{\dagger}-a_{-}a_{+}a_{+}^{\dagger}a_{-}^{\dagger}-a_{-}a_{+}a_{-}^{\dagger}a_{+}^{\dagger}-a_{-}a_{-}^{\dagger}a_{+}a_{+}^{\dagger}-a_{-}a_{-}^{\dagger}a_{-}a_{-}^{\dagger}+a_{-}^{2}\left(a_{-}^{\dagger}\right)^{2}\right\} ,
\end{array}\label{eq:54}
\end{equation}
then, we obtain
\begin{equation}
\left\langle 0,0\right|\mathscr{T}_{4}\left|0,0\right\rangle =\frac{1}{4}\frac{\hbar^{2}}{m}\tau.\label{eq:55}
\end{equation}

Finally, using Eqs. (\ref{eq:49}), (\ref{eq:51}), (\ref{eq:53}),
and (\ref{eq:55}), the energy shift for the ground state, representing
the first-order correction to the energy eigenvalues, is given by
\begin{equation}
E_{0}^{(1)}=\triangle E=\frac{1}{2}\frac{\hbar^{2}}{m}\tau.\label{eq:56}
\end{equation}

Thus, the complete expression for the energy levels of the deformed
Schrödinger oscillator, incorporating the first-order correction,
is
\begin{equation}
E_{n,l}^{\textrm{DNC}}=E_{n,l}^{\textrm{NC}}+\triangle E=\hbar\omega n-\frac{1}{2}m\omega^{2}l\Theta+\frac{1}{2}\frac{\hbar^{2}}{m}\tau,\label{eq:57}
\end{equation}
which encapsulates the contributions from both DNC and NC spaces.
Setting $\tau=\Theta=0$, retrieves the well-established results of
the commutative space. It is important to note that the second-order
correction to the ground state energy $E_{0}^{(2)}$, is given by
\begin{equation}
E_{0}^{(2)}=\frac{\left|\left\langle 1,0\right|\mathscr{H}^{(\tau)}\left|0,0\right\rangle \right|^{2}}{E_{0}^{(0)}-E_{1,0}^{(0)}}+\frac{\left|\left\langle 0,1\right|\mathscr{H}^{(\tau)}\left|0,0\right\rangle \right|^{2}}{E_{0}^{(0)}-E_{0,1}^{(0)}}.\label{eq:58}
\end{equation}

Here, $E_{0}^{(0)}$ represents the unperturbed energy of the ground
state, while $E_{1,0}^{(0)}$ and $E_{0,1}^{(0)}$ correspond to the
unperturbed energy levels of the excited states $\mathrm{\mid1,0>}$
and $\mathrm{\mid0,1>},$ respectively. It should be emphasized that
the first excited state is two-fold degenerate $\mathrm{\left|1,0\right\rangle }$
and $\left|0,1\right\rangle $. The associated perturbed $\mathrm{2\times2}$
matrix is given by:
\begin{equation}
\left[\begin{array}{cc}
\left\langle 0,1\right|\mathscr{H}^{(\tau)}\left|0,1\right\rangle  & \left\langle 0,1\right|\mathscr{H}^{(\tau)}\left|1,0\right\rangle \\
\left\langle 1,0\right|\mathscr{H}^{(\tau)}\left|0,1\right\rangle  & \left\langle 1,0\right|\mathscr{H}^{(\tau)}\left|1,0\right\rangle 
\end{array}\right].\label{eq:59}
\end{equation}

By applying the same method, we can determine the second-order correction
to the ground-state energy of the deformed Schrödinger oscillator. 

We remark that the energy (\ref{eq:57}) is real, so that this is
an answer to the question of Bender and Boettcher \citep{key-29}.
Furthermore, as the DNC parameter vanishes $\tau=0$, we obtain the
same result as found in the NC space (without the NCy of momentum)
in \citep{key-3}. 

\begin{figure}[h]
\centering{}%
\begin{tabular}{cc}
\includegraphics[scale=0.58]{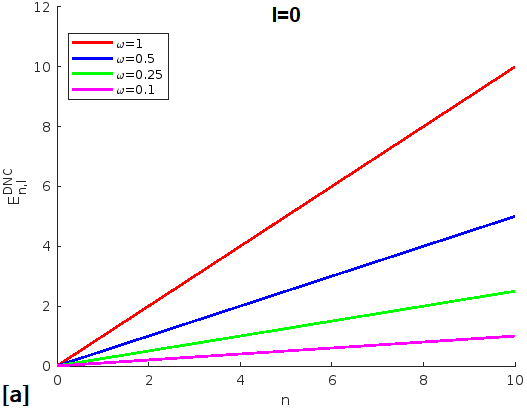} & \includegraphics[scale=0.58]{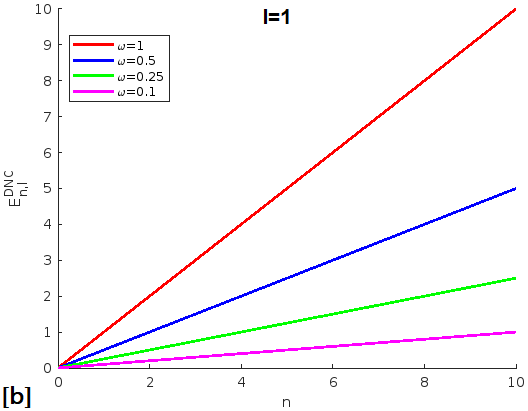}\tabularnewline
\end{tabular}\caption{\label{3}The DNC energy eigenvalues $E_{n,l}^{\textrm{DNC}}$ as
a function of the quantum number $n$, shown for different values
of $\omega$, with fixed $\Theta=\tau=0.001$. (a) $l=0$ ; (b) $l=1$.}
\end{figure}

\begin{figure}[h]
\centering{}%
\begin{tabular}{cc}
\includegraphics[scale=0.58]{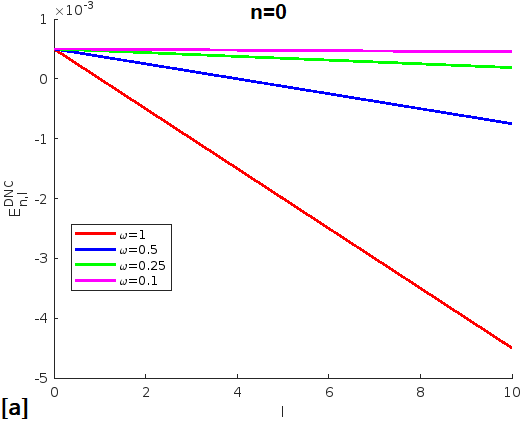} & \includegraphics[scale=0.58]{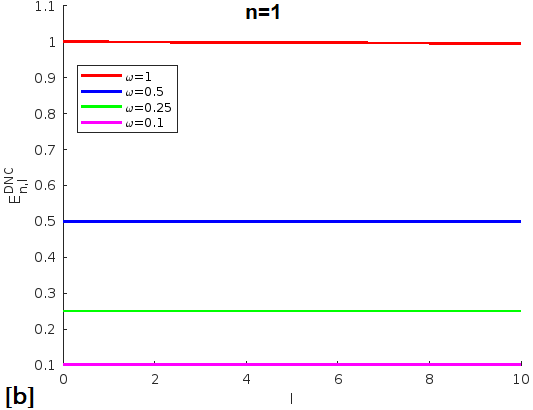}\tabularnewline
\end{tabular}\caption{\label{4}The DNC energy eigenvalues $E_{n,l}^{\textrm{DNC}}$ as
a function of the quantum number $l$, shown for different values
of $\omega$, with fixed $\Theta=\tau=0.001$. (a) $n=0$ ; (b) $n=1$.}
\end{figure}

\begin{figure}[h]
\centering{}%
\begin{tabular}{cc}
\includegraphics[scale=0.58]{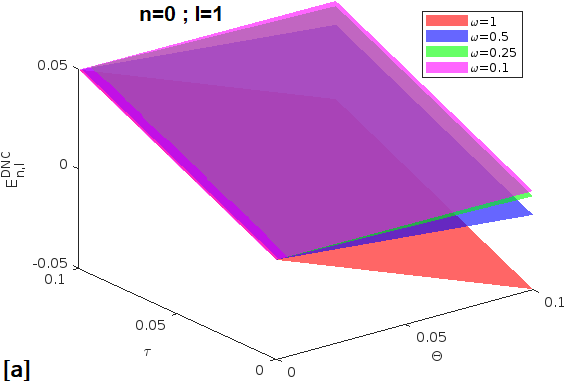} & \includegraphics[scale=0.58]{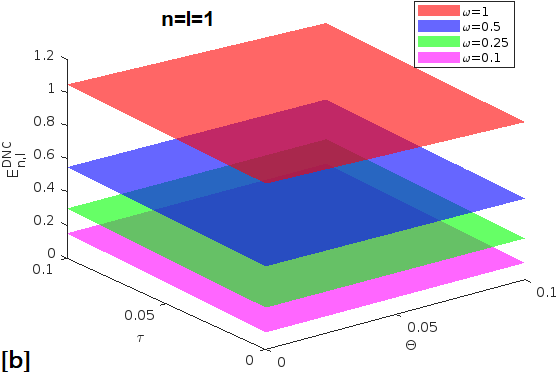}\tabularnewline
\end{tabular}\caption{\label{5}3D plots of the NC energy eigenvalues $E_{n,l}^{\textrm{NC}}$
as a function of NC and DNC parameters $\Theta$ and $\tau$, shown
for different values of $\omega$. (a) $n=0$, $l=1$ ; (b) $n=l=1$.}
\end{figure}

In Figs. \ref{3}, \ref{4}, and \ref{5}, we present the spectrum
of the system as a function of the deformation parameters $\tau$
and $\Theta$ for different oscillator angular frequency $\omega$
and selected quantum numbers where the constants are set to $\hbar=m=1$.
The results show that the effects of the deformation parameters on
the energy levels are significant for typical values of $\omega$.

\section{{\normalsize\label{IV}}Thermal properties of the DNC Schrödinger
oscillator}

In the present analysis, the thermodynamic properties are investigated
within the conventional canonical ensemble formalism. Although NC
and DNC geometries are motivated by ideas originating from string
theory and quantum gravity, the present treatment is intended as an
effective description of the modified quantum dynamics rather than
a formulation of fundamental quantum-gravitational thermodynamics.
Consequently, the effects of NCy are incorporated through the corrected
energy spectrum, while the standard Boltzmann--Gibbs factor $\textrm{e}^{-\beta E}$
is assumed to remain valid for the equilibrium statistical description
of the system.

\subsection{{\normalsize\label{IV-A}}Partition function and thermal quantities}

In this part, we explore the influence of DNC and NC spaces on the
thermal properties of the Schrödinger oscillator. Within the framework
of statistical physics, these properties can be derived from the partition
function $Z\left(\beta\right)$. Therefore, we begin by computing
the partition function for the deformed Schrödinger oscillator, which
is obtained by summing over all possible energy levels of the system
(single oscillator) as follows:
\begin{equation}
Z\left(\beta\right)=\sum_{n=0}^{\infty}\textrm{e}^{-\beta E_{n}}\:\textrm{ with }\beta=\frac{1}{K_{B}T},\label{eq:63}
\end{equation}
where $\beta$ is the Boltzmann factor, $E_{n}$ represents the energy
spectrum, $T$ denotes the absolute temperature, and $K_{B}$ is the
Boltzmann constant. Now, based on the form of the energy eigenvalues
(\ref{eq:57}), the single-oscillator partition function is written
as
\begin{equation}
Z\left(\beta,\Theta,\tau\right)=\sum_{n=0}^{\infty}Q^{n}\textrm{e}^{\frac{\beta}{2}\left(m\omega^{2}\Theta l-\frac{\tau}{m}\hbar^{2}\right)},\text{ with }Q=\textrm{e}^{-\beta\hbar\omega}.\label{eq:64}
\end{equation}

In the present analysis, we consider a fixed value of the angular
quantum number $l$ (taken as $l=l_{0}$) in order to isolate the
effects of the deformation parameters, in particular the contribution
of the NC parameter $\Theta$ on the thermodynamic behavior of the
system. Since $Q<1$,  we have
\begin{equation}
\sum_{n=0}^{\infty}Q^{n}=\frac{1}{1-Q},\label{eq:65}
\end{equation}
hence, the partition function within DNC space becomes
\begin{equation}
Z\left(\beta,\Theta,\tau\right)=\frac{\textrm{e}^{\frac{\beta}{2}\left(m\omega^{2}\Theta l-\frac{\tau}{m}\hbar^{2}\right)}}{1-\textrm{e}^{-\beta\hbar\omega}}.\label{eq:66}
\end{equation}

Note that for a set of N-particles, we have
\begin{equation}
\mathcal{Z}\left(\beta,\Theta,\tau,N\right)=Z^{N}\left(\beta,\Theta,\tau\right)=\frac{\textrm{e}^{N\beta\left(\frac{1}{2}m\omega^{2}\Theta l-\frac{\tau}{2m}\hbar^{2}\right)}}{\left(1-\textrm{e}^{-\beta\hbar\omega}\right)^{N}}.\label{eq:67}
\end{equation}

According to Eq. (\ref{eq:66}), we can see that the DNC partition
function, $Z\left(\beta,\Theta,\tau\right)$, depends on the DNC and
NC parameters. Key thermal proprieties such as free energy $F$, internal
energy $U$, entropy $S$, and specific heat capacity $C$ can be
systematically computed with the help of the following relations:
\begin{equation}
\begin{array}{l}
F=-\frac{1}{\beta}\ln Z,\\
U=-\frac{\partial\ln Z}{\partial\beta},\\
S=\ln Z-\beta\frac{\partial\ln Z}{\partial\beta},\\
C=\beta^{2}\frac{\partial^{2}\ln Z}{\partial\beta^{2}}.
\end{array}\label{eq:68}
\end{equation}

We start from Eq. (\ref{eq:66}) giving
\begin{equation}
\ln Z=\frac{\beta}{2}\frac{\hbar^{2}}{m}\tau-\frac{\beta}{2}m\omega^{2}l\Theta-\ln\left(1-\textrm{e}^{-\beta\hbar\omega}\right),\label{eq:70}
\end{equation}
and
\begin{equation}
\frac{\partial\ln Z}{\partial\beta}=\frac{\hbar^{2}}{2m}\tau-\frac{1}{2}m\omega^{2}l\Theta-\hbar\omega\frac{\textrm{e}^{-\beta\hbar\omega}}{1-\textrm{e}^{-\beta\hbar\omega}},\label{eq:71}
\end{equation}
and
\begin{equation}
\frac{\partial^{2}\ln Z}{\partial\beta^{2}}=\left\{ \frac{\hbar\omega}{2\sinh\left(\beta\hbar\omega/2\right)}\right\} ^{2}.\label{eq:72}
\end{equation}

Consequently, the thermal quantities defined by Eq. (\ref{eq:68})
become
\begin{equation}
\begin{array}{l}
F=\frac{1}{2}m\omega^{2}l\Theta-\frac{\hbar^{2}}{2m}\tau+\frac{1}{\beta}\ln\left(1-\textrm{e}^{-\beta\hbar\omega}\right),\\
U=\frac{1}{2}m\omega^{2}l\Theta-\frac{\hbar^{2}}{2m}\tau+\hbar\omega\frac{\textrm{e}^{-\beta\hbar\omega}}{1-\textrm{e}^{-\beta\hbar\omega}},\\
S=\beta\hbar\omega\frac{\textrm{e}^{-\beta\hbar\omega}}{1-\textrm{e}^{-\beta\hbar\omega}}-\ln\left(1-\textrm{e}^{-\beta\hbar\omega}\right),\\
C=\left\{ \frac{\hbar\omega\beta}{2\sinh\left(\beta\hbar\omega/2\right)}\right\} ^{2}.
\end{array}\label{eq:73}
\end{equation}

It is evident that both NC and DNC spaces influence the thermal quantities---except
for the entropy and the specific heat capacity, which do not exhibit
such effects. 

It is also important to point out that when the deformed Schrödinger
oscillator system is subjected to a uniform magnetic field $\overrightarrow{B}$,
the magnetic behavior of it can be systematically explored by analyzing
the magnetization $M$, and magnetic susceptibility $\chi$, i.e.,
\citep{key-30,key-31} $M=\frac{1}{\beta}\frac{1}{Z}\frac{\partial\ln Z}{\partial\overrightarrow{B}}$,
and $\chi=\frac{\partial M}{\partial\overrightarrow{B}}.$ The magnetization
is often used to examine the possibility of phase transitions, while
magnetic susceptibility quantifies how much a material becomes magnetized
under an applied magnetic field. Susceptibility also helps clarify
the nature of a phase transition, if it exists. 

Next, we thoroughly investigate the behavior of the derived thermal
quantities, presenting the numerical results and illustrating the
various effects considered on these quantities.

\subsection{{\normalsize\label{IV-B}}Results and discussion}

\begin{figure}[H]
\centering{}%
\begin{tabular}{cc}
\includegraphics[scale=0.6]{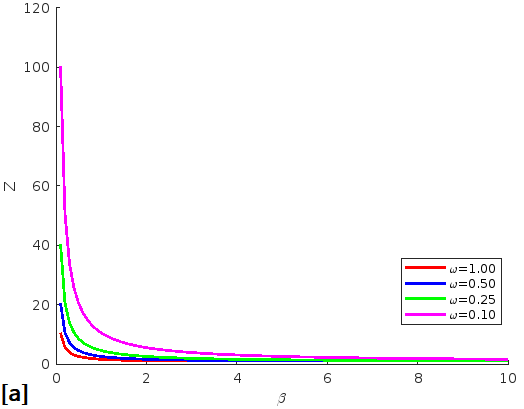} & \includegraphics[scale=0.6]{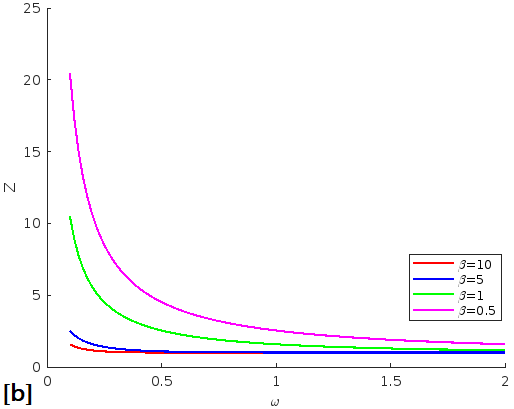}\tabularnewline
\end{tabular}\caption{\label{6}The partition function $Z$ as a function of $\beta$ and
$\omega$ separately, with fixed values o f $\Theta=\tau=0.001$.
(a) Variation of $Z$ versus $\beta$ for different values of $\omega$;
(b) variation of $Z$ versus $\omega$ for different values of $\beta$.}
\end{figure}

\begin{figure}[H]
\centering{}%
\begin{tabular}{cc}
\includegraphics[scale=0.6]{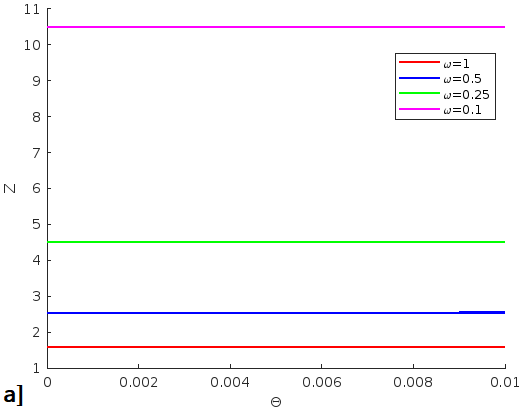} & \includegraphics[scale=0.6]{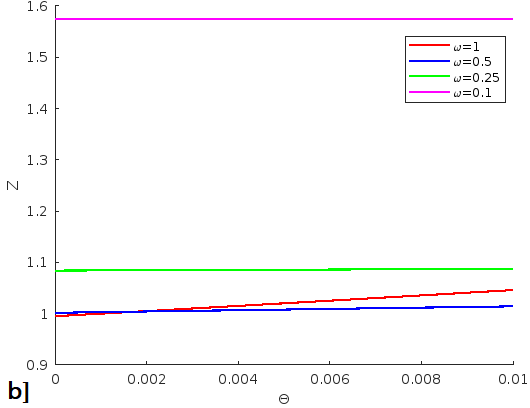}\tabularnewline
\includegraphics[scale=0.6]{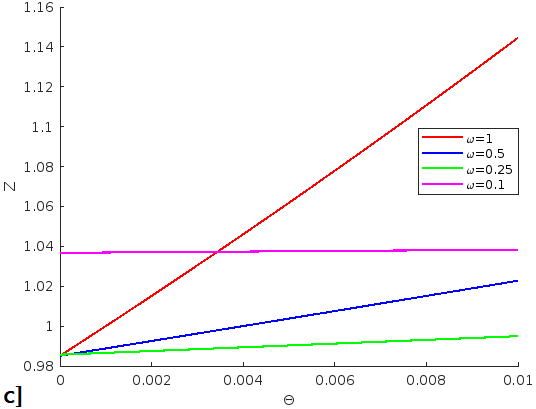} & \includegraphics[scale=0.6]{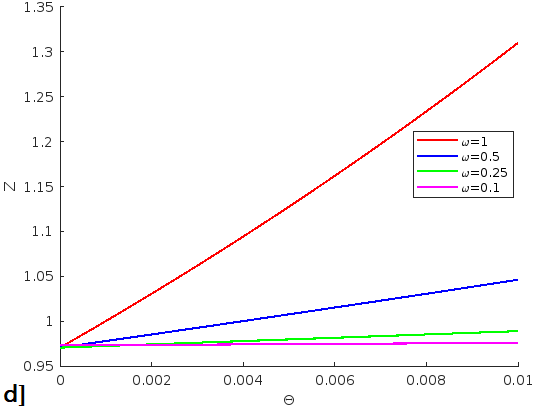}\tabularnewline
\end{tabular}\caption{\label{7}The partition function $Z$ as a function of the NC parameter
$\Theta$, shown for different values of $\omega$, with $\tau=0.01$
fixed. (a) $\beta=1$; (b) $\beta=10$; (c) $\beta=30$; (d) $\beta=60$.}
\end{figure}

\begin{figure}[H]
\centering{}%
\begin{tabular}{cc}
\includegraphics[scale=0.6]{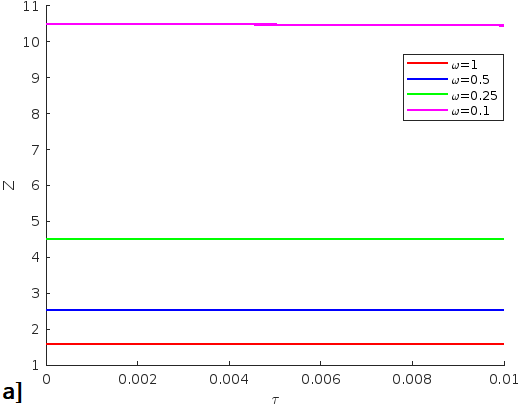} & \includegraphics[scale=0.6]{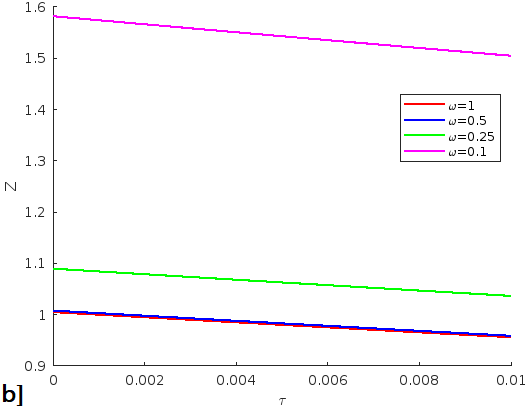}\tabularnewline
\includegraphics[scale=0.6]{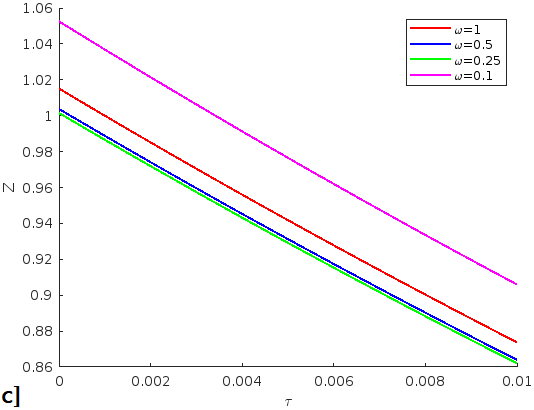} & \includegraphics[scale=0.6]{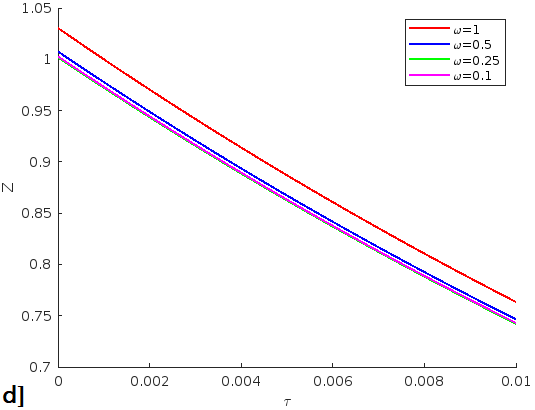}\tabularnewline
\end{tabular}\caption{\label{8}The partition function $Z$ as a function of the DNC parameter
$\tau$, shown for different values of $\omega$, with $\Theta=0.01$
fixed. (a) $\beta=1$; (b) $\beta=10$; (c) $\beta=30$; (d) $\beta=60$.}
\end{figure}

\begin{figure}[H]
\centering{}%
\begin{tabular}{cc}
\includegraphics[scale=0.6]{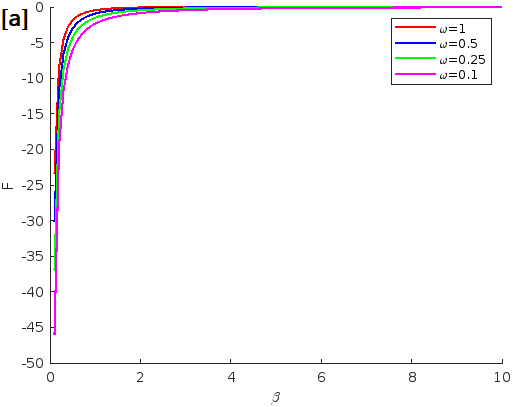} & \includegraphics[scale=0.6]{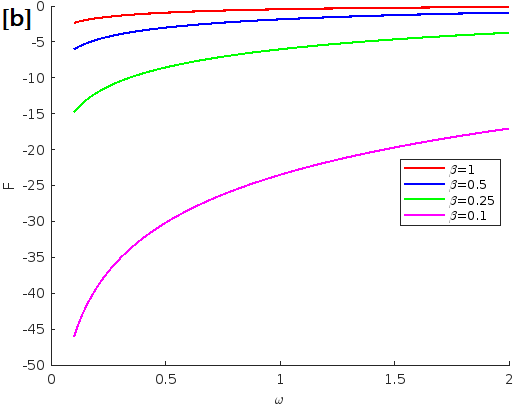}\tabularnewline
\end{tabular}\caption{\label{9}The free energy $F$ as a function of $\beta$ and $\omega$
separately, with fixed values of $\Theta=\tau=0.001$. (a) Variation
of $F$ with respect to $\beta$ for different values of $\omega$;
(b) variation of $F$ with respect to $\omega$ for different values
of $\beta$.}
\end{figure}

\begin{figure}[H]
\centering{}%
\begin{tabular}{cc}
\includegraphics[scale=0.6]{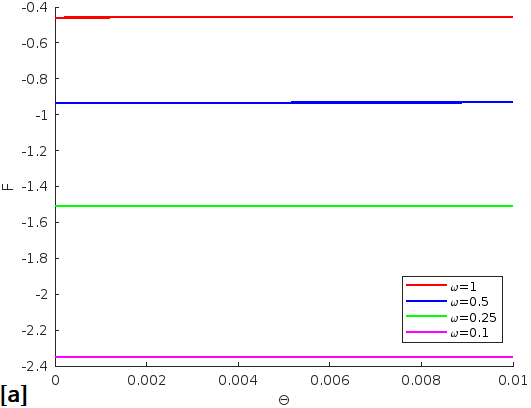} & \includegraphics[scale=0.6]{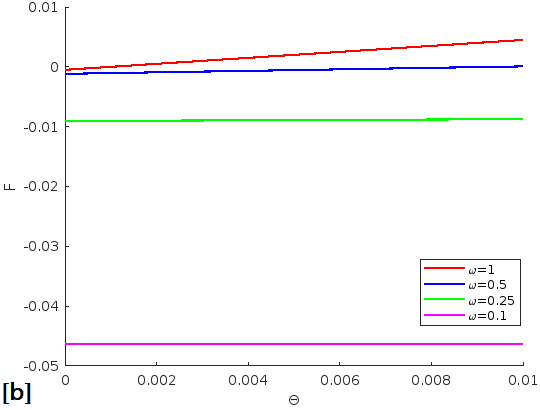}\tabularnewline
\includegraphics[scale=0.6]{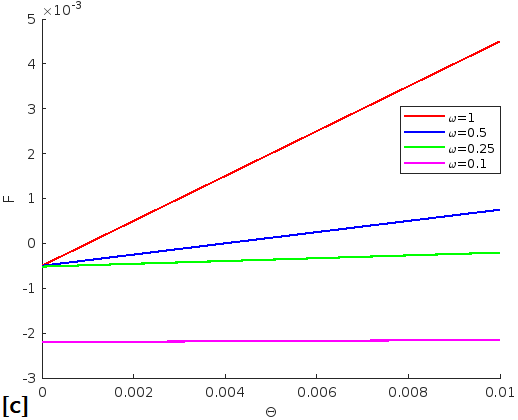} & \includegraphics[scale=0.6]{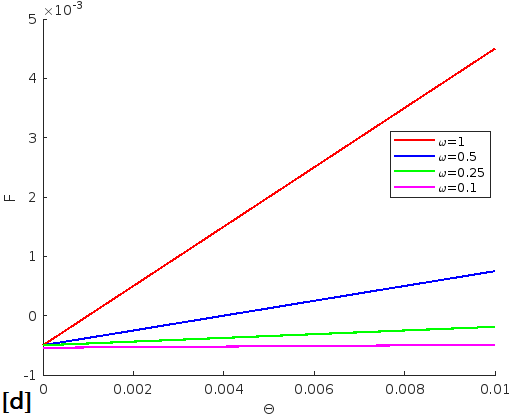}\tabularnewline
\end{tabular}\caption{\label{10}The free energy $F$ as a function of $\Theta$, shown
for different values of $\omega$, with $\tau=0.001$ fixed. (a) $\beta=1$;
(b) $\beta=10$; (c) $\beta=30$; (d) $\beta=60$.}
\end{figure}

\begin{figure}[H]
\centering{}%
\begin{tabular}{cc}
\includegraphics[scale=0.58]{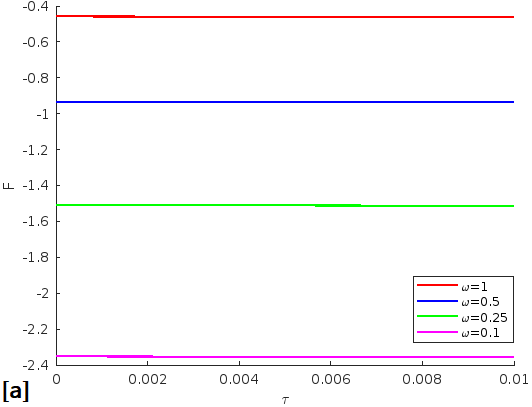} & \includegraphics[scale=0.58]{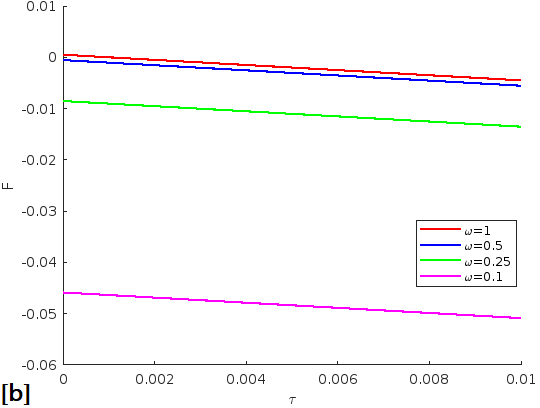}\tabularnewline
\includegraphics[scale=0.58]{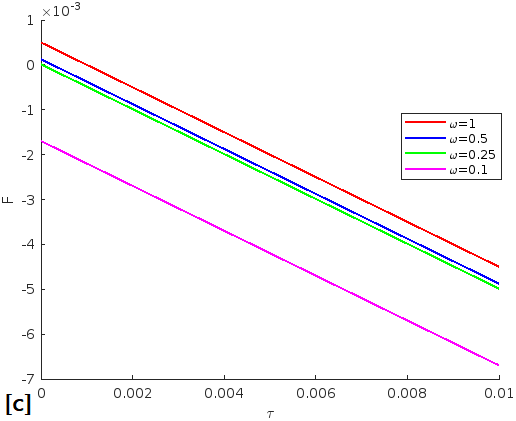} & \includegraphics[scale=0.58]{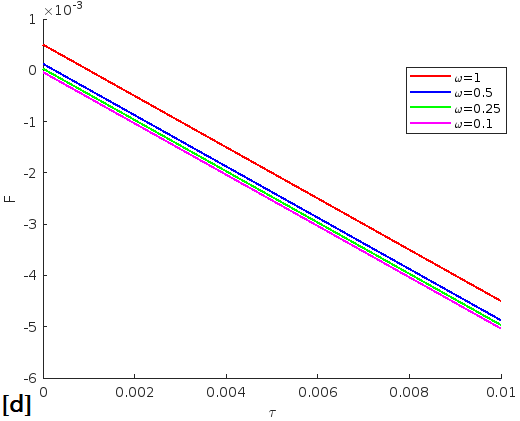}\tabularnewline
\end{tabular}\caption{\label{11}The free energy $F$ as a function of $\tau$, shown for
different values of $\omega$, with $\Theta=0.001$ fixed. (a) $\beta=1$;
(b) $\beta=10$; (c) $\beta=30$; (d) $\beta=60$.}
\end{figure}

\begin{figure}[H]
\centering{}%
\begin{tabular}{cc}
\includegraphics[scale=0.6]{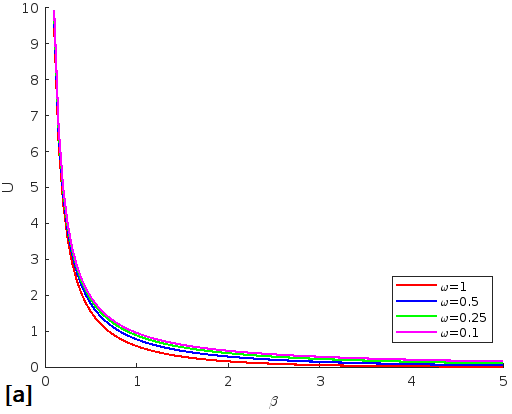} & \includegraphics[scale=0.6]{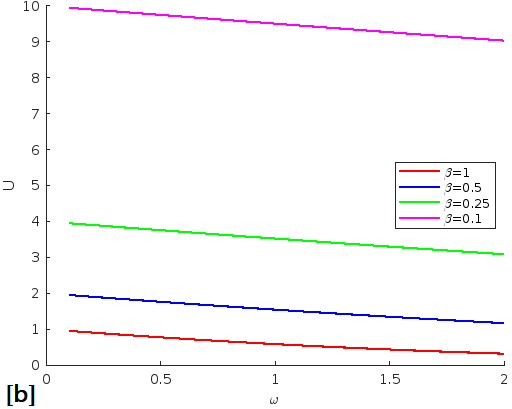}\tabularnewline
\end{tabular}\caption{\label{12}The internal energy $U$ as a function of $\beta$ and
$\omega$ separately, with fixed values o f $\Theta=\tau=0.001$.
(a) Variation of $U$ with respect to $\beta$ for different values
of $\omega$; (b) variation of $U$ with respect to $\omega$ for
different values of $\beta$.}
\end{figure}

\begin{figure}[H]
\centering{}%
\begin{tabular}{cc}
\includegraphics[scale=0.6]{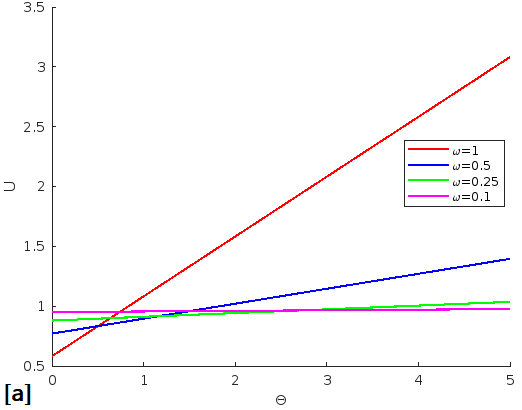} & \includegraphics[scale=0.6]{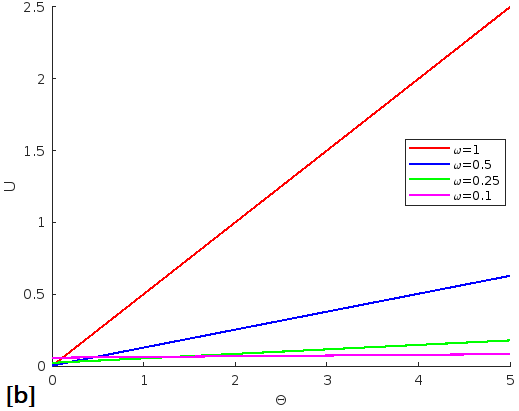}\tabularnewline
\includegraphics[scale=0.6]{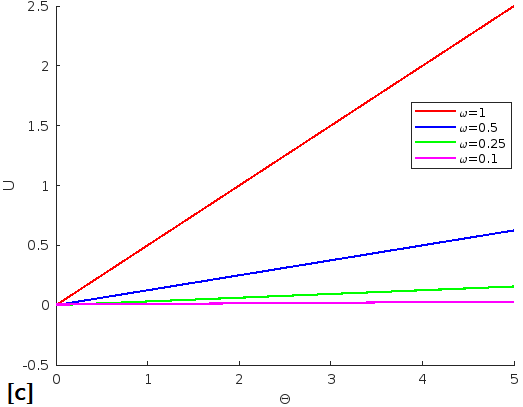} & \includegraphics[scale=0.6]{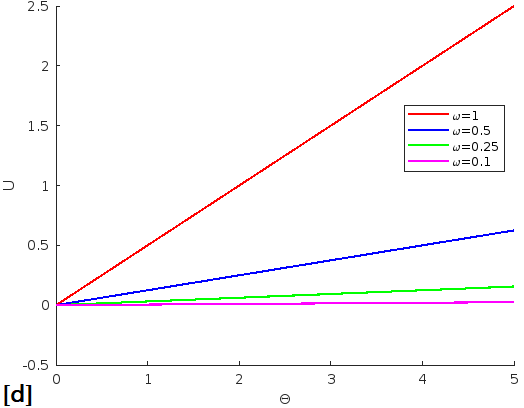}\tabularnewline
\end{tabular}\caption{\label{13}The internal energy $U$ as a function of $\Theta,$shown
for different values of $\omega$, with $\tau=0.001$ fixed. (a) $\beta=1$;
(b) $\beta=10$; (c) $\beta=30$; (d) $\beta=60$.}
\end{figure}

\begin{figure}[H]
\centering{}%
\begin{tabular}{cc}
\includegraphics[scale=0.6]{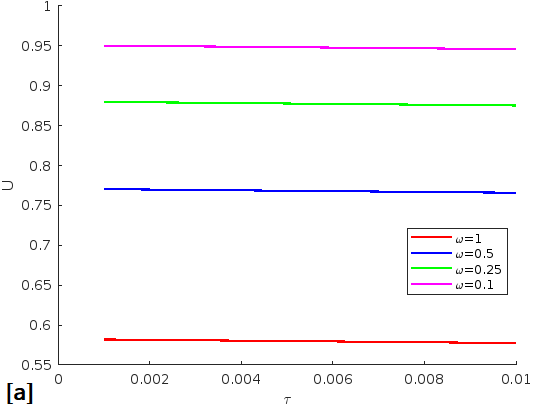} & \includegraphics[scale=0.6]{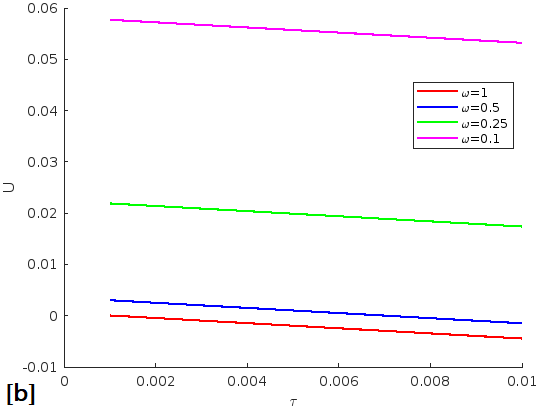}\tabularnewline
\includegraphics[scale=0.6]{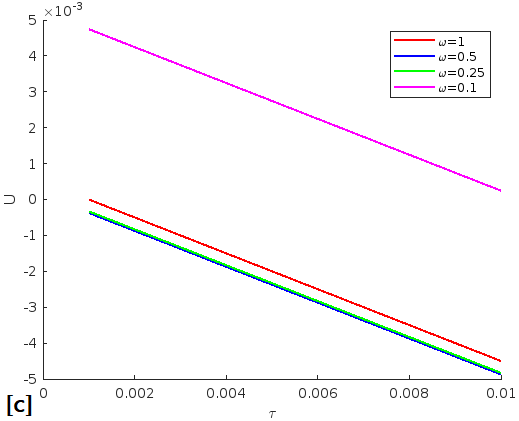} & \includegraphics[scale=0.6]{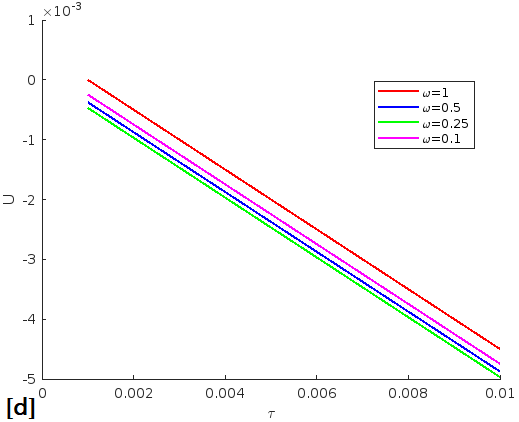}\tabularnewline
\end{tabular}\caption{\label{14}The free energy $U$ as a function of $\tau$, shown for
different values of $\omega$, with $\Theta=0.001$ fixed. (a) $\beta=1$;
(b) $\beta=10$; (c) $\beta=30$; (d) $\beta=60$.}
\end{figure}

\begin{figure}[H]
\centering{}%
\begin{tabular}{cc}
\includegraphics[scale=0.6]{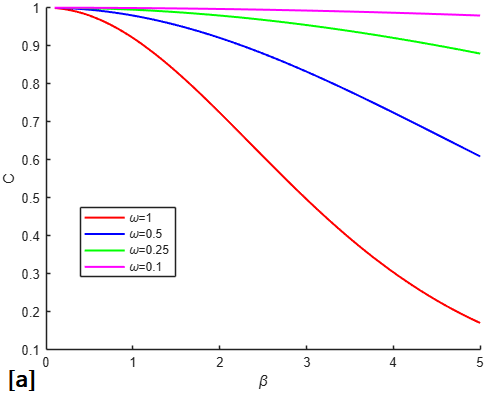} & \includegraphics[scale=0.6]{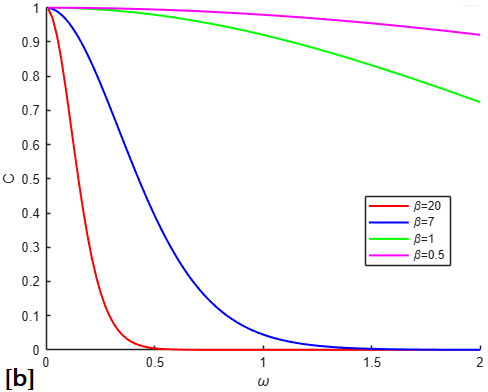}\tabularnewline
\end{tabular}\caption{\label{15}The specific heat capacity $C$ as a function of $\beta$
and $\omega$ separately. (a) Variation of $C$ with respect to $\beta$
for different values of $\omega$; (b) variation of $C$ with respect
to $\omega$ for different values of $\beta$.}
\end{figure}

\begin{figure}[H]
\centering{}%
\begin{tabular}{cc}
\includegraphics[scale=0.6]{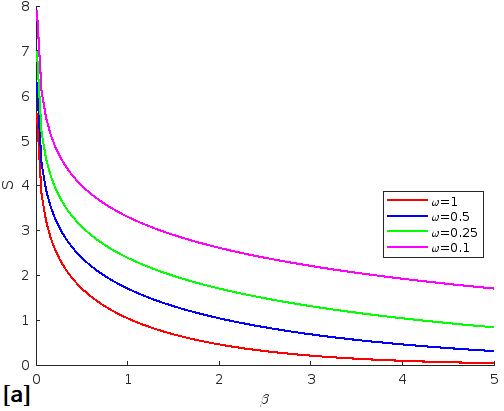} & \includegraphics[scale=0.6]{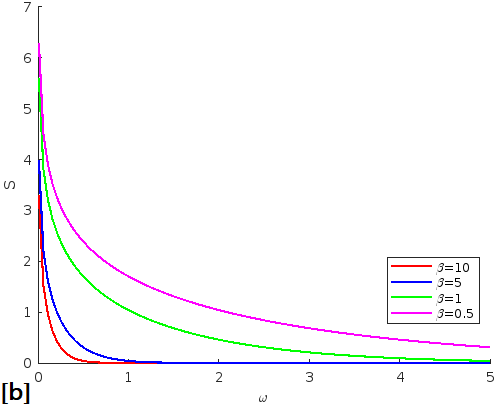}\tabularnewline
\end{tabular}\caption{\label{16}The entropy $S$ as a function of $\beta$ and $\omega$
separately. (a) Variation of $S$ with respect to $\beta$ for different
values of $\omega$; (b) variation of $S$ with respect to $\omega$
for different values of $\beta$.}
\end{figure}

We present numerical results for the thermal properties of the deformed
Schrödinger oscillator and its partition function under varying thermal
($\beta$), quantum ($\omega$), and deformation ($\Theta$,$\tau$)
parameters. The constants are set to $\hbar=m=l=1$. The partition
function (Eq. (\ref{eq:66})) is illustrated in Figs. \ref{6}--\ref{8}.
The thermal quantities (Eq. (\ref{eq:73})) are depicted in Figs.
\ref{9}--\ref{16}.

It should be noted that $\beta$ (the inverse temperature) determines
the system\textquoteright s regime through the dimensionless parameter
$\beta\hbar\omega$. This parameter indicates whether thermal fluctuations
dominate ($\beta\hbar\omega\ll1$) or quantum effects prevail ($\beta\hbar\omega\gg1$).
The system's behavior is dictated by the interplay between thermal
and quantum fluctuations.
\begin{itemize}
\item At $\beta=0.1$ (high temperatures), $\beta\hbar\omega<1$; thus,
the classical behavior prevails due to strong thermal fluctuations. 
\item For $\beta=1$ (moderate temperature), $\beta\hbar\omega\sim\omega$;
thus, thermal and quantum contributions compete comparably.
\item For $\beta=10$ (low temperatures), $\beta\hbar\omega>1$; thus, quantum
fluctuations dominate, and the system favors low-energy states.
\item At $\beta=100$ (ultra-low temperatures), $\beta\hbar\omega\gg1$;
thus, the oscillator exhibits near-pure quantum behavior, and the
thermal noise is nearly absent.
\end{itemize}
The practical implications on thermal quantities vary significantly
across the values of $\beta$. These regimes determine whether the
thermal behavior aligns with classical predictions or exhibits quantum
fluctuations. It should be emphasized that Eq. (\ref{eq:66}) converges
for all $\beta>0$ (i.e., all finite temperatures), ensuring its validity
across all thermal regimes. Furthermore, the values used of $\beta$
are employed in simulations and experiments to investigate for instance
phase transitions, quantum-to-classical crossovers, and low-temperature
phenomena. Based on this distinction, one can classify the following
cases, assuming $\omega=1$: 

$\phantom{}$
\begin{center}
\begin{tabular}{lV{\linewidth}ll>{\raggedright}p{8.5cm}}
\hline 
{\small$\beta$} & {\small Regime} & {\small$\beta\hbar\omega$} & {\small Dominant effects} & {\small Experimental relevance}\tabularnewline
\hline 
{\small 0.1} & {\small High temperature} & {\small$\textrm{<1}$} & {\small Thermal fluctuations} & {\small Classical limit validation (e.g., room-temperature magnetism)}\tabularnewline
{\small 1} & {\small Quantum-classical}{\small\par}

{\small crossover} & {\small$\textrm{\ensuremath{\sim}1}$} & {\small Competing effects} & {\small Phase transition studies (e.g., quantum critical materials)}\tabularnewline
{\small 10} & {\small Low temperature} & {\small$\textrm{>1}$} & {\small Quantum fluctuations} & {\small Quantum technologies (e.g., qubits, sensors) }\tabularnewline
{\small 100} & {\small Ultra-low temperature}{\small\par}

{\small (Millikelvin regime)} & {\small$\textrm{\ensuremath{\gg}1}$} & {\small Near-pure quantum} & {\small Quantum spin liquids, topological order (e.g., millikelvin
experiments, superconducting circuits)}\tabularnewline
\hline 
\end{tabular}
\par\end{center}

Alternatively, quantum fluctuations are significant when $k_{B}T\ll\hbar\omega$,
whereas thermal fluctuations dominate when $k_{B}T\gg\hbar\omega$.
Additionally, adjusting $\omega$ shifts the crossover scale $\beta\hbar\omega$.

We plot the partition function $Z$, free energy $F$, and internal
energy $U$ as functions of the parameters $\beta$ and $\omega$
separately, with $\Theta$ and $\tau$ held fixed, in Figs. \ref{6},
\ref{9}, and \ref{12}, respectively. 

In Figs. \ref{7}, \ref{10}, and \ref{13}, the same quantities---$Z$,
$F$, and $U$---are plotted versus $\Theta$, while in Figs. \ref{8},
\ref{11}, and \ref{14}, they are plotted versus $\tau$, all for
a fixed $\beta$ and shown for different values of $\omega$. Finally,
Figs. \ref{15} and \ref{16} extend the analysis by illustrating
separately the dependence of the specific heat capacity $C$ and entropy
$S$ on $\beta$ and $\omega$, respectively.

$\phantom{}$

In Fig. \ref{6}, the partition function $Z$ is plotted as a function
of $\beta$ and $\omega$ separately, with fixed $\Theta$ and $\tau$.
Figs. \ref{7} and \ref{8} extend this analysis, showing $Z$\textquoteright s
dependence on $\Theta$ and $\tau$, respectively, across varying
$\omega$ and $\beta$. The results demonstrate distinct behaviors:
increasing $\Theta$ elevates $Z$, while increasing $\tau$ suppresses
it, reflecting their opposing roles in the deformed energy corrections.
The deformation parameters significantly influence $Z$, and the effects
become significant as $\beta$ increases. Specifically, $Z$ decreases
with $\beta$ (lower temperatures) due to thermal suppression, but
this trend is modulated by $\omega$, which amplifies the oscillator\textquoteright s
energy contribution. The interplay is most pronounced at intermediate
$\beta$, where $\Theta$-driven enhancements counteract thermal suppression,
whereas $\tau$-induced reductions further dampen $Z$. These deviations
underscore how NC geometry reshapes thermodynamics, with $\Theta$
and $\tau$ acting as competing parameters in the partition function\textquoteright s
behavior.  

We plot the free energy $F$ as a function of the parameters $\beta$
and $\omega$ separately, with $\Theta$ and $\tau$ fixed in Fig.
\ref{9}. In Figs.  \ref{10} and \ref{11}, we depict $F$ as a function
of the deformation parameters $\Theta$ and $\tau$, respectively,
for different values of $\omega$ and across various cases of $\beta$.
It is shown that the effect on $F$ becomes significant as $\beta$
increases ({\small low temperatur}es). Additionally, there are distinct
behaviors: increasing $\Theta$ elevates $F$, while increasing $\tau$
decreases it, reflecting their opposing roles in the partition function
and energy corrections.

In Figure \ref{12}, we plot the internal energy $U$ as a function
of the parameters $\beta$ and $\omega$ separately, with $\Theta$
and $\tau$ fixed. It is shown that exhibits contrasting trends: it
diminishes with $\beta$ due to suppressed thermal excitations related
to the quantum statistical term $\textrm{\ensuremath{\exp}[-\ensuremath{\beta\hbar\omega}]}$
at low temperatures but grows quadratically with $\omega$, particularly
at high $\beta$, reflecting the dominance of the $\omega^{2}\Theta$
term. Figure \ref{13} highlights $U$\textquoteright s linear rise
with $\Theta$, amplified by larger $\omega$. Figure \ref{14} reveals
$U$\textquoteright s reduction with $\tau$, intensified at high
$\beta$, aligning with the $\tau$-dependent term\textquoteright s
subtractive role. Together, the figures underscore the interplay between
quantum statistics (governed by $\beta$) and deformation parameters:
$\Theta$ enhances energy contributions, while $\tau$ suppresses
them, with both effects magnified in low-temperature (high-$\beta$)
regimes.

In Figure \ref{15}, we depict the specific heat $C$ as a function
of the parameters $\beta$ and $\omega$ separately. The results show
that $C$ decreases rapidly with increasing $\beta$ (corresponding
to the low-temperature regime), which is consistent with the $\sinh$-dominated
suppression appearing in its analytical expression. Furthermore, the
dependence of $C$ on $\omega$ reflects a competition between the
quadratic growth in $\omega$ and the exponential suppression induced
by the hyperbolic sine term. In Figure \ref{16}, we plot the entropy
$S$ as a function of the parameters $\beta$ and $\omega$ separately.
Here, $S$ decreasing with $\beta$, approaching zero at low temperatures
(high $\beta$), as expected from the logarithmic and Bose-like terms.
The $\omega$-dependence of $S$ highlights reduced entropy for larger
$\omega$ due to quantized energy spacing (quantized energy spacing
restricts accessible states), aligning with quantum statistical trends.
Both figures emphasize the interplay between thermal ($\beta$) and
quantum ($\hbar\omega$) scales in governing thermodynamic behavior.

\section{{\normalsize\label{V}}Conclusion and remarks}

In this study, we have analytically investigated the effects of NC
and DNC spaces on the 2D Schrödinger oscillator using perturbation
theory. The energy eigenvalues were derived, and the influence of
NC and DNC parameters was thoroughly examined. In particular, we considered
the first-order correction to the energy for the ground state, revealing
that the energy levels explicitly depend on the parameters $\Theta$
and $\tau$, as shown in Eq. (\ref{eq:57}). To further analyze these
effects, we presented graphical representations of the energy spectrum
under various conditions, allowing for a detailed exploration of the
system's behavior. Note that by employing a 2D linear Bopp-shift,
we successfully mapped the deformed Schrödinger oscillator system
to its commutative counterpart.  Furthermore, building on our work
in \citep{key-16}, we derived the nonrelativistic limit of the Klein-Gordon
oscillator in DNC space using the mapping $(E^{2}-m^{2}c^{4})\rightarrow2mc^{2}\epsilon$,
where $\epsilon$ represents the nonrelativistic energy. This procedure
exactly reproduces the DNC Schrödinger oscillator, yielding the same
deformed Hamiltonians (Eqs. (\ref{eq:26}--\ref{eq:28}) and eigensystem
(Eq. (\ref{eq:57})), as well as the identical energy shift (Eq. (\ref{eq:56})).
Given this complete correspondence, our findings strongly support
the conclusion that the nonrelativistic limit of the Klein-Gordon
equation leads to the Schrödinger oscillator rather than the quantum
harmonic oscillator. 

From a physical perspective, the dependence of the energy spectrum
on the deformation parameters $\Theta$ and $\tau$ shows that NCy
acts as an effective deformation of the underlying phase-space, leading
to shifts in the energy levels that could, in principle, carry phenomenological
significance. Although such effects are expected to be extremely small
at accessible energy scales, they provide a useful theoretical framework
for exploring possible signatures of NC and DNC structures. In addition,
the position-dependent nature of DNC space introduces an effective
minimal length scale, which may be relevant in the context of quantum
gravity and string-inspired models where short-distance structure
plays a fundamental role. This suggests that DNC geometry encodes
effective geometric properties of space rather than directly observable
fundamental constants. Moreover, the deformation affects the energy
spectrum of the system and consequently influences thermodynamic quantities
such as the internal energy and specific heat, thereby providing insight
into how NC effects may manifest in statistical and thermal properties
of quantum systems. Finally, the presence of non-Hermitian features
in the DNC framework raises further questions regarding the role of
Hermiticity and possible \ensuremath{\mathscr{P}}\ensuremath{\mathscr{T}}-symmetric
extensions, indicating that such deformed structures may open new
directions for studying consistent quantum dynamics in generalized
phase-space geometries. The significance of investigating DNC spaces
stems from their fundamental connection to string theory, where objects
naturally exhibit extended structures. Notably, in the limit $\tau\rightarrow0$,
our results recover those of NC quantum mechanics (as discussed in
\citep{key-3}), while for $\Theta\rightarrow0$, the system reduces
to standard quantum mechanics. This confirms the consistency and compatibility
of our findings. Additionally, we explored the thermal properties
of the system by computing the partition function. The obtained results
were visualized across multiple scenarios, providing deeper insights
into the thermal response of the deformed system. 

Overall, our study opens avenues for further investigations into non-Hermitian
position-dependent algebras and deformed quantum systems. As emphasized
in Section \ref{II}, operators in DNC space exhibit intrinsic non-Hermitian
characteristics. Therefore, the interplay between DNC geometry, non-Hermitian
quantum mechanics, and string theory constitutes a promising direction
for uncovering novel physical phenomena.

\section*{Funding}

This research received no external funding. 

\section*{Declarations Conflict of interest}

The author declares no conflict of interest.

\section*{Data availability}

The datasets used and analysed during the current study are available
from the corresponding author on reasonable request.

\end{document}